\documentclass[a4paper,10pt]{article}
\usepackage[utf8]{inputenc}
\usepackage[fleqn,cmex10]{amsmath}
\usepackage{booktabs}
\usepackage{color,colortbl}  
\usepackage{amssymb}
\usepackage{pbox}
\usepackage{verbatim}
\usepackage{natbib}
\usepackage{graphicx}
\usepackage[margin=1.2in]{geometry}

\definecolor{Blue}{rgb}{0,0,1}
\definecolor{darkgreen}{rgb}{0,0.35,0}
\definecolor{Gray}{gray}{0.75}
\definecolor{LightCyan}{rgb}{0.80,1,1}

\newenvironment{mysplit}%
  {\arraycolsep 0pt \begin{array}{l}}%
  {\end{array}}

\title{Probabilistic Complex Event Recognition: A Survey}
\author{E. Alevizos$^1$, A. Skarlatidis$^1$, \\ A. Artikis$^2$,$^1$, G. Paliouras$^1$\\
\textit{$^1$NCSR Demokritos, Athens, Greece}\\
\textit{$^2$University of Piraeus, Greece}}
\date{}

\begin{document}

\maketitle

\begin{abstract}
Complex Event Recognition applications exhibit various types of uncertainty, 
ranging from incomplete and erroneous data streams to imperfect complex event patterns. 
We review Complex Event Recognition techniques that handle, to some extent, uncertainty. 
We examine techniques based on automata, probabilistic graphical models and first-order logic, which are the most common ones, 
and approaches based on Petri Nets and Grammars,
which are less frequently used. 
A number of limitations are identified with respect to the employed languages,  
their probabilistic models and their performance, as compared to the purely deterministic cases.
Based on those limitations,
we highlight promising directions for future work.
\end{abstract}

\section{Introduction}
Systems for Complex Event Recognition 
accept as input a stream 
of time-stamped, simple, derived events (SDE)s. 
A SDE (``low-level event'') is the result of applying a computational derivation 
process to some other event, such as an event coming from a sensor. 
Using SDEs as input, CER systems identify complex events (CE)s of interest--
collections of events that satisfy some pattern. 
The ``definition'' of a CE (``high-level event'') imposes temporal and, possibly, 
atemporal constraints on its sub-events (SDEs or other CEs). 
Consider, for example, the recognition of attacks on computer network nodes, 
given the TCP/IP messages.
A CER system attempting to detect a Denial of Service attack has to identify (as one possible scenario)
both a forged IP address that fails to respond and that the rate of requests is unusually high.  
In maritime monitoring,
in order to detect an instance of illegal fishing,
a CER system has to perform both some geospatial tasks,
such as estimating whether a vessel is moving inside a protected area,
and temporal ones,
like determining whether a vessel spent a significant amount of time in this area.  
In this sense, 
CER (event pattern matching) is one of the functionalities of Complex Event Processing (CEP).

The SDEs arriving at a CER
system almost always carry a certain degree of uncertainty and/or ambiguity.
Information sources might be heterogeneous, with data of different schemas,
they might fail to respond or even send corrupt data. 
Even if we assume perfectly accurate sensors, the domain under study might be
difficult or impossible to model precisely, 
thereby leading to another type of uncertainty. 
Until recently, most CER systems did not make any effort to handle uncertainty
(it is instructive to see the relevant discussion about uncertainty in \cite{cugola_processing_2011}). 
This need is gradually being acknowledged and it seems that this might
constitute a major line of research and development for CER.

The purpose of this paper is to present an overview of existing approaches for
CER under uncertainty.
Since this field is relatively new,
without a substantial number of contributions coming from researchers directly
involved with CER,
we have chosen to adopt a broader perspective 
and include methods targeting activity recognition
and scene understanding on image sequences coming from video sources.
Although activity recognition is a field  with its own requirements,
it is related closely enough to CER so that some of the ideas
and methods applied there might provide inspiration to CER researchers as well.
However, it is not our intention to present a survey of video recognition methods
and we have selectively chosen those among them that we believe are closer to
CER
(for a survey of activity recognition methods from video sources, 
see \cite{vishwakarma_survey_2013}).
We have used two basic criteria for our choice (applied to the CER methods as well).
First, we require that the method employs some kind of relational formalism to
describe activities, 
since purely propositional approaches are not sufficient for CER.
Second, we require that uncertainty be handled within a probabilistic framework,
since this is a framework that provides clear and formal semantics.
In this respect, 
our work is related to previously conducted comparisons within
the field of statistical relational learning, 
both theoretical \cite{de_raedt_probabilistic_2003,jaeger_model-theoretic_2008,muggleton_behavioral_2008}
and practical \cite{bruynooghe_exercise_2009}.
The related field of query processing over uncertain data in
probabilistic databases/streams is covered in other surveys 
(e.g., \cite{wang_survey_2013}) and, therefore, 
we will not include such papers in our survey.
 
Throughout the remainder of the paper, we are going to use a running example in
order to assess the presented approaches against a common domain.
Our example comes from the domain of
video recognition. 
We assume that a CER engine receives as input a set of time-stamped
events, 
derived from cameras recording a basketball game.
However, we need to stress that our input events are not composed of raw
images/frames and that the task of the CER engine is not to perform
image processing. 
We assume availability of algorithms that can perform the corresponding tasks, 
such as object recognition and tracking.
Therefore, our SDEs consist of events referring to objects and
persons, like \textit{walking}, \textit{running} and \textit{dribbling}.
The purpose is to define patterns for the recognition of some high-level
long-term activities, 
e.g., that a \textit{matchup} between two players or a \textit{double-teaming} is
taking place.

The structure of the paper is as follows:
Section \ref{section:framework} discusses briefly the types of uncertainty that
may be encountered in a CER application.
In Section \ref{section:scope} we present certain criteria based on which 
we included (or excluded) papers from our survey and 
in Section \ref{section:eval} we discuss the dimensions along which a proposed 
solution for handling uncertainty may be evaluated.
Section \ref{section:approaches} presents the reviewed approaches. 
Finally, Section \ref{section:conclusions} summarizes them in a tabular form and comments on their limitations.
Some open issues and lines of potential future work are also identified.

\section{Uncertainty in Event Recognition}
\label{section:framework}
Understanding uncertainty in its different types is
crucial for any CER system that aspires to provide an efficient way of handling
it.
The ideal CER system would be capable of handling
all types of uncertainty within a unified and mathematically rigorous
framework. 
However, this is not always possible and the current CER systems are still far
from achieving such an ideal.
Different domains might be susceptible to different types of uncertainty,
while different CER engines employ various methods for responding to
it, 
ranging from the ones that simply ignore it to those that use highly complex, fully-fledged,
dynamic, probabilistic networks. 
In this section, we give a brief description and classification of
the various types of uncertainty that may be encountered by a CER system.
For further discussion,
see \cite{wasserkrug_taxonomy_2006,artikis_event_2012}. 

\subsection{Data Uncertainty}
\label{data_uncertainty}
The event streams that provide the input data to a CER engine can exhibit
various types of uncertainty. 
In this section, we present the main types of uncertainty that may be found in
incoming event streams.

One type of uncertainty is that of incomplete or missing evidence.
A sensor may fail to report certain events, 
for example due to some hardware malfunction.
Even if the hardware infrastructure works as expected, 
certain characteristics of the monitored environment could prevent events from
being recorded, 
e.g. an occluded object in video monitoring or a voice being drowned 
by stronger acoustic signals.  

The events of the input stream may have a noise component added to them.
In this case, events may be accompanied by a probability value. 
There are many factors which can contribute to the corruption of the input
stream, such as the limited accuracy of sensors or distortion along a
communication channel.
Another distinction which might be important in certain contexts is that
between stochastic and systematic noise, 
e.g. the video frames from a camera may exhibit a systematic noise component, 
due to different light conditions throughout the day.

When noise corrupts the input event stream, 
a CER system might find itself in a position where it receives events asserting
contradictory statements. 
For example, in a computer vision application which needs to track objects, 
such as that of our running example,
if there are multiple software classifiers, one of them 
may assert the presence of an object (e.g., the ball) whereas another may indicate that
no such object has been detected.

Finally, when a CER system needs to learn the structure and the
parameters of a probabilistic model from training data, 
quite often the data are inconsistently annotated.
Therefore, the rules to be learned have to incorporate this uncertainty and
carry a confidence value.

\subsection{Pattern Uncertainty}
\label{rules_uncertainty}
Besides the uncertainty present in the input data, 
a noise-tolerant CER system should also be able to handle cases where
the event patterns are not precise or complete.

Due to lack of knowledge or due to the inherent complexity of a domain, 
it is sometimes impossible to capture exactly all the conditions that a pattern
should satisfy. 
It might also be preferable and less costly to provide a more general
definition of a pattern which is easy to implement rather than trying to exactly
determine all of its conditions.
A pattern with a wider scope, 
which does not have to check multiple conditions,
may also be more efficient to compute and,
in some cases,
this performance gain could be more critical than accuracy. 
In both cases, we cannot infer an event with certainty and a mechanism is required
to quantify our confidence.

For example, 
a rule for determining when a team is attempting an offensive move 
might be defined as a pattern in which one of the team's players has the ball
and all other players are located in the opponent team's half-court.
However, 
the same pattern could also be satisfied when a player is attempting a free throw
or for an out-of-bounds play.
Depending on our requirements,
we might or might not want to include all of these instances as cases of offensive moves.
Defining all of these sub-cases would require more refined conditions,
something which is not always possible.
Yet, we might be still interested in capturing this pattern and provide a confidence value.

\section{Scope of the survey}
\label{section:scope}

Before presenting our framework and evaluation dimensions, 
we explain the rationale behind our choices
and clarify some basic conceptual issues, 
which we deem important from the point of view of CER. 

\subsection{Probabilistic models}
A seemingly simple method to handle uncertainty is to ignore or remove noise through pre-processing
or filtering of the data, thus facilitating the use of a deterministic model. 
Other methods are available as well, 
such as possibilistic reasoning, 
conflict resolution (accept data according to the trustworthiness of a source)
and fuzzy sets.
For example, a logic-based method is proposed by Shet et al \citeyear{ShetNRD07,ShetSBRND11},
which employs logic programming and the Bilattice
framework \cite{ginsberg1988multivalued}.
Another example is the work presented in \cite{ma_event_2010},
where the Dempster-Shafer theory is used in order to take into account
the trustworthiness of heterogeneous event sources.
We focus on probabilistic models because
they provide a unified and rigorous framework
and the bulk of research on CER under uncertainty employs such models.

\subsection{Time representation}
Some approaches, especially those employing dynamic graphical models,
resort to an implicit representation of time,
whereby time slices depend on (some of) the previous slice(s),
without taking into account time itself as a variable.
Useful as this solution might be in domains characterized by sequential
patterns, 
such as activity recognition in video,
there are other cases in CER where time constraints need to be explicitly
stated.
Although we include in our survey some approaches with an implicit time representation, 
our focus will mostly be on methods with explicit time
representation.

\subsection{Relational models}
A substantial proportion of the existing probabilistic models are propositional by nature,
as is the case with many probabilistic graphical models,
such as simple Bayesian networks.
Probabilistic graphical models have been successfully applied to a
variety of CER tasks where a significant amount of uncertainty
exists.
Especially within the machine vision community, they seem to be one of the most
frequently used approaches.
Since CER requires the processing of streams of time-stamped
SDEs, numerous CER methods are based on sequential variants of
probabilistic graphical models, 
such as Hidden Markov Models \cite{rabiner1986HMM} and their extensions
(e.g., coupled \cite{brand_coupled_1997},
Dynamically Multi-Linked \cite{gong_recognition_2003}
and logical Hidden Markov Models \cite{kersting2006lohmm}), 
Dynamic Bayesian Networks \cite{murphy2002DBN} and 
Conditional Random Fields \cite{LaffertyMP01CRF}.

As far as Hidden Markov Models are concerned,
since they are generative models, 
they require an elaborate process of extracting the correct independence
assumptions and they perform inference on the complete set of possible worlds.
Moreover, their first-order nature imposes independence assumptions
with regard to the temporal sequence of events 
(with only the current and the immediately previous states taken into account)
that might not be realistic for all domains.
On the other hand, 
Conditional Random Fields are discriminative models, a feature which allows them to avoid the
explicit specification of all dependencies and, as a consequence,
avoid imposing non-realistic independence assumptions \cite{vail_conditional_2007,wu_joint_2007,liao_hierarchical_2005}.
However, 
both Hidden Markov Models and Conditional Random Fields assume a static domain of objects (with the exception of
logical Hidden Markov Models \cite{kersting2006lohmm}), 
whereas a CER engine cannot make the same assumption,
since it is not possible to determine beforehand all the possible
objects that might appear in an input stream from a dynamic and evolving environment. 
Additionally, the lack of a formal representation language makes the definition of
structured CEs (Complex Events) complicated and the use of background knowledge very hard.
From a CER perspective, these issues constitute a severe limitation,
since rules for detecting CEs often require relational and hierarchical structures,
with complex temporal and atemporal relationships.
For these reasons,
we do not discuss Hidden Markov Models and Conditional Random Fields in a more detailed manner. 
Instead, we focus our investigation on methods with relational models.

\section{Evaluation Dimensions}
\label{section:eval}

In this section, 
we provide a general framework for the
discussion of the different approaches and establish a number of evaluation
dimensions against which the strengths and weaknesses of each method may be assessed.
We follow the customary division between representation and inference.
As far as learning is concerned, 
although in general it is a very active research area within the statistical relational learning community,
we have decided not to include a detailed discussion about the learning
capabilities of the examined approaches in our survey. 
The reason is that very few of the probabilistic CER systems deal with learning
(these exceptions are mentioned in Section \ref{section:approaches}).
Instead, we will try to draw some conclusions about the
performance of each system, 
taking into account the difficulties in making performance comparisons.

\subsection{Representation}
\subsubsection*{A simple unifying event algebra}
\label{section:eval:rep}
We begin our discussion of representation by introducing a basic notation
for CER.
For a more detailed discussion of the theory behind CER, 
we refer readers to \cite{luckham_power_2001,etzion_action_2010,cugola_tesla_2010}.
Following the terminology of \cite{luckham_power_2001}, 
we define an event as an object in the form of a tuple of data components,
signifying an activity and holding certain relationships to other events by
time, causality and aggregation. 
An event with $N$ attributes can be represented as 
$\mathit{EventType}(Attr_{1}, \ldots, Attr_{N}, Time)$,
where $Time$ might be a point, in case of an instantaneous event, 
or an interval during which the event happens, if it is durative.
Notice, however, that, when timepoints are used,
some unintended semantics might be introduced,
as discussed in \cite{paschke_ecaruleml_2006}.
For our running example,
events could be of the form $\mathit{EventType}(\mathit{PlayerName}, \mathit{UnixTime})$
and one such an event could be the following:
$\mathit{Running}(Antetokounmpo, 19873294673)$.
In CER, 
we are interested in detecting patterns of events among the streams of SDEs (Simple Derived Events).
Therefore, we need a language for expressing such pattern detection rules.
For example, 
by using '$;$' as the sequence operator,
the pattern
\begin{equation*}
\begin{aligned}
& EventType_{1}(A^{1}_{1},\ldots,A^{1}_{N},T_{1});\\ 
& EventType_{2}(A^{2}_{1},\ldots,A^{2}_{N},T_{2})
\end{aligned}
\end{equation*}
would serve to detect instances where an event of type $Type_{1}$ is followed by an
event of type $Type_{2}$.
An example would be:
\begin{equation*}
\begin{aligned}
& Running(Antetokounmpo, 19873294673);\\ 
& Jumping(Antetokounmpo, 19873294677);\\
& Dunking(Antetokounmpo, 19873294680)
\end{aligned}
\end{equation*}

Based on the capabilities of existing CER systems and probabilistic CER methods, 
we adopt here a simple event algebra. 
Formalisms for reasoning about events and time have appeared in the past, 
such as the Event Calculus \cite{kowalski_logicbased_1986,cervesato_calculus_2000}
and Allen's Interval Algebra \cite{allen_maintaining_1983,allen_towards_1984},
and have already been used for defining event algebras
(e.g., \cite{paschke_knowledge_2008}).
With the help of the theory of descriptive complexity, 
recent work has also identified those constructs of an event algebra which
strike a balance between expressive power and complexity \cite{zhang_complexity_2014}. 
Our event algebra will be defined in a fashion similar to the above mentioned
efforts, borrowing mostly from \cite{zhang_complexity_2014,cervesato_calculus_2000}.

Below, we present the syntax of the event algebra:
\begin{equation*}
\begin{aligned}
ce ::= \ & \ sde\ & | \ &  	 \\
\ & ce_{1}\ ;\ ce_{2}\ & | \ & Sequence\\
\ & ce_{1}\ \vee \ ce_{2}\ & | \ & Disjunction \\
\ & ce^{*}\ & | \ & Iteration \\
\ & \neg\ ce\ & | \ & Negation \\
\ & \sigma_{\theta}(ce) \ & | \ & Selection \\
\ & \pi_{m}(ce) \ & | \ & Production \\
\ & [ce]_{T_{1}}^{T_{2}} & \ & Windowing \ \text{(from $T_{1}$ to $T_{2}$)}
\end{aligned} 
\end{equation*}
where $\sigma_{\theta}(ce(v_{1},\ldots,v_{n}))$ \textit{selects} those $ce$
whose variables $v_{i}$ satisfy the set of predicates $\theta$ and
$\pi_{m}(ce(a_{1},\ldots,a_{n}))$ \textit{returns} a $ce$ whose attribute values are
a possibly transformed subset of the attribute values of $a_{i}$ of the initial
$ce$, according to a set of mapping expressions $m$. 
$Conjunction$ may be written as 
$ce ::= ce_{1} \wedge ce_{2} ::= (ce_{1};ce_{2}) \vee (ce_{2};ce_{1})$.

The following list explains the above operations:
\begin{itemize}
  \item $Sequence$: Two events following each other in time.
  \item $Disjunction$: Either of two events occurring, regardless of temporal relations.
        Conjunction (both events occurring) may be expressed by combining 
        $Sequence$ and $Disjunction$. 
  \item $Iteration$: An event occurring $N$ times in sequence, 
        where $N \geq 0$.
  \item $Negation$: Absence of event occurrence.
  \item $Selection$: Select those events whose attributes satisfy a set of 
        predicates/relations, temporal or otherwise.
  \item $Production$: Return an event whose attribute values are a 
        possibly transformed subset of the attribute values of its sub-events.
  \item $Windowing$: Evaluate the conditions of an event pattern within a specified time window.
\end{itemize}

For example, 
a traveling violation,
occurring when a player who has possession of the ball takes more than two steps without dribbling,
could be defined as follows:
\begin{equation*}
\begin{aligned}
 & traveling(P^{'},T^{'})::= \pi_{P^{'}=P_{3},T^{'}=T_{3}}(\sigma_{P_{1}=P_{2},P_{2}=P_{3}}( \\
 & (hasBall(P_{1},T_{1}) \wedge takesStep(P_{1},T_{1}) \wedge \neg dribbling(P_{1},T_{1}) ) \ ; & \\
 & (hasBall(P_{2},T_{2}) \wedge takesStep(P_{2},T_{2}) \wedge \neg dribbling(P_{2},T_{2}) ) \ ; & \\
 & (hasBall(P_{3},T_{3}) \wedge takesStep(P_{3},T_{3}) \wedge \neg dribbling(P_{3},T_{3}) ) \ )) & \\
\end{aligned}
\end{equation*}

The above algebra is simple, but expressive,
defining temporal constraints between events.
For example, in the above rule about $\mathit{traveling}$,
the $\mathit{sequence}$ operator ($;$) implies that 
$T_{2} > T_{1}$ and $T_{3} > T_{2}$.  
Note that some CER systems (e.g., the Chronicle Recognition System 
\cite{dousson_situation_1993,dousson_extending_2002,dousson_chronicle_2007}) 
allow the predicates $\theta$ to 
be applied directly to the attribute of time.
Throughout the remainder of the paper, 
we also adopt the selection strategy of $skip-till-any-match$
(irrelevant events are ignored and relevant events can satisfy multiple
``instances'' of a rule) and the $zero-consumption$ strategy 
(an event can trigger multiple rules). 

The above syntax allows for the construction of event hierarchies, 
a crucial capability for every CER system.
Being able to define events at various levels and reuse those intermediate
inferred events in order to infer other, higher-level events is not trivial.
Theoretically, every event language could achieve this simply by embedding the
patterns of lower-level events into those at higher levels,
wherever they are needed. 
However, this solution would result in long and contrived patterns and would
incur heavy performance costs, since intermediate events would need to
be computed multiple times.
Moreover, there are multiple ways a system could handle the propagation of
probabilities from low-level to high-level events and these differences can
affect both the performance and the accuracy of the system.

\subsubsection*{Probabilistic Data}
The event algebra defined above is deterministic.
We now extend it in order to take uncertainty into account. 
As we have already discussed, 
we can have uncertainty both in the data and the patterns. 
As far as data uncertainty is concerned,
we might be uncertain about both the occurrence of an event and about the
values of its attributes.
For example, the ProbLog2 system \cite{fierens_inference_2013} employs annotated disjunctions.
Therefore, for a probabilistic event, 
we could write
$Prob::EventType(Value_{1}, \ldots, Value_{N}, Time)$, 
which means that this event with these values for its attributes might have
occurred with probability $Prob$ and not have occurred at all with probability
$1-Prob$.
In order to assign probabilities to attribute values,
e.g. for two different values of $Attribute_{1}$, 
we could write
\begin{equation*}
\begin{aligned}
& Prob_{1}:: \ & EventType(Value_{1}^{1}, \ldots, Value_{N}, Time) \ & ; \\
& Prob_{2}:: \ & EventType(Value_{1}^{2}, \ldots, Value_{N}, Time) \ &
\end{aligned}
\end{equation*}
In this case, the sum of the two probabilities should not exceed the value of
$1$ and, 
if it is below $1$, 
this means that there is also a probability of no occurrence, 
whose value is $1-\sum_{s}Prob_{s}$.
We assume that probabilistic events are represented as discrete random variables.

With respect to the probability space,
a common assumption is that it is defined over the possible histories of the
probabilistic SDEs. 
If SDEs are defined as discrete random variables,
then one SDE history corresponds to making a choice about each of the SDEs
among mutually exclusive alternative choices.
The probability distribution is then defined over those SDE histories.
E.g., if we have the following probabilistic SDEs
\begin{equation*}
\begin{aligned}
0.8:: Running(Antetokounmpo, 19873294673)\\ 
0.6:: Jumping(Antetokounmpo, 19873294677)\\
0.7:: Dunking(Antetokounmpo, 19873294680)
\end{aligned}
\end{equation*}
then the probability space is composed of the 8 possible histories obtained 
through all the combinations of event (non-)occurrence.
Choosing the history in which all events do occur would yield a probability of $0.8*0.6*0.7$,
assuming that all SDEs are independent (which is not always the case).

\subsubsection*{Probabilistic Model}
In addition to handling uncertain data, 
we also require probabilistic rules.
We express a probabilistic rule by appending its
probability value as a prefix, e.g.
\begin{equation*} 
Prob::ce(A,T)::=\pi_{A=A_{2},T=T_{2}}(ce_{1}(A_{1},T_{1});ce_{2}(A_{2},T_{2}))
\end{equation*}
where, if $ce_{2}$ occurred after $ce_{1}$,
then $ce$ occurred at $T_{2}$ with probability $\mathit{Prob}$. 
The probability space is extended to include the inferred CE in the event histories.
A probabilistic rule should then be understood as defining the conditional
probability of the CE occurring,
given that its sub-events occurred and satisfied its pattern.
The attribute values of this CE are those returned by the $production$ operator $\pi$.
If we need to define different probability values to the 
CE with different attribute values,
we could use again the syntax of annotated disjunctions, e.g.
\begin{equation*}
\begin{aligned}
& Prob:: \ ce(T,A)::= \\
& \pi_{T=T_{2},A=A_{2}}(ce_{1}(T_{1},A_{1});ce_{2}(T_{2},A_{2})) \ & ; \\
& Prob^{'}:: \ ce(T,A)::= \\
& \pi_{T=T_{2},A=2*A_{2}}(ce_{1}(T_{1},A_{1});ce_{2}(T_{2},A_{2})) \ &
\end{aligned}
\end{equation*}
where the occurrence probability of $ce$,
with its attribute $A$ having value $A_{2}$,
is $\mathit{Prob}$,
whereas it may also have occurred with $A=2*A_{2}$
and probability $\mathit{Prob^{'}}$.  

There are other ways to define the probability space and its semantics.
For example, in the probabilistic programming literature it is common to  
use the possible worlds semantics for the probability space 
(e.g., in ProbLog \cite{fierens_inference_2013}).
The probability distribution is defined over the (possibly multi-valued)
Herbrand interpretations of the theory, as encoded by the CE patterns.
In this setting, we could assign non-zero probabilities even in cases where the
rule is violated and we could end up with every Herbrand interpretation being a
model/possible world. 
The existence of ``hard'' rules which must be satisfied excludes
certain interpretations from being considered as models.
When using grammars (and sometimes logic),
the space might be defined over the possible proofs that lead to the
recognition/acceptance of a CE.

\subsection{Inference}

In probabilistic CER, the task is often to compute the marginal probabilities of the CEs,
given the evidence SDEs.
Consider the following example:
\begin{equation*}
P(\mathit{offense}(\mathit{MilwaukeeBucks},[00:00,00:24]) | \mathit{SDEs})
\end{equation*}
where we want to calculate the probability that the team of $\mathit{MilwaukeeBucks}$
was on the offensive for the first $24$ seconds of the game and we assume that
$\mathit{offense}(\mathit{team},[\mathit{start},\mathit{end}])$ is a durative CE, 
defined over intervals and in terms of SDEs, 
such as $\mathit{running}$, $\mathit{dribbling}$, etc. 
Moreover, there are cases when we might be interested in performing maximum a
posteriori (MAP) inference, in which the task is to compute the most probable
states of some CEs, given the evidence SDE stream.
A simple example from basketball is the query
about the most probable time interval during which an offense by a team is
taking place:
\begin{equation*}
\begin{aligned}
I_{\mathit{offense}} = \underset{I}{arg}maxP(\mathit{offense}(\mathit{MilwaukeeBucks},I) | \mathit{SDEs}) \ &
\end{aligned}
\end{equation*}

Another dimension concerns the ability to perform approximate inference. 
For all but the simplest cases, 
exact inference stumbles upon serious performance issues,
unless several simplifying assumptions are made.
For this reason, approximate inference is considered essential.
It is also possible for a system to provide answers with confidence intervals and/or the option
of setting a confidence threshold above which an answer may be accepted.

\subsection{Performance}

CER systems are usually evaluated for their performance in terms of throughput,
measured as the number of events processed per second 
and latency, as measured by the average time required to process an event. 
Less often, the memory footprint is reported. 
Reporting throughput figures is not enough by itself, 
since there are multiple factors which can affect performance.
For example, the number of different event types in the SDE stream or the rule
selectivity, i.e. the percentage of SDEs selected by a rule, are such factors
(see \cite{mendes_performance_2009} for a comprehensive list of performance
affecting factors). 
When uncertainty is introduced, 
the complexity of the problem increases and other factors that affect performance 
enter the picture, such as the option of approximate inference.
Unfortunately, standard benchmarks specifically targeting probabilistic CER have not yet been established.
Therefore, in this survey we can report only what is available in the examined papers.

Systems need to be evaluated along another dimension as well,
that of accuracy.
Precision and recall are the usual measures of accuracy. 
F-measure is the harmonic mean of precision and recall.
The issue of accuracy is of critical importance and is not orthogonal
to that of performance.
A system may choose to sacrifice accuracy in favor of performance by adopting
techniques for approximate inference.
Another option would be to make certain simplifying assumptions with respect to
the dependency relationships between events so that the probability space
remains tractable.

\section{Approaches}
\label{section:approaches}

Surprisingly, there haven't been many research efforts devoted
exclusively to the problem of handling uncertainty 
within the community of distributed event-based systems.
The majority of research papers that could be deemed as relevant to our problem
actually come from the computer vision community.
Perhaps it is not much of a surprise if one takes into account the historical
roots of CER systems.
Stemming from the need to build more active databases and to operate upon
streams of data that 
have a pre-defined schema, 
the problem of uncertainty, 
although present, 
was not as critical as in the case of efficiently processing events from sensors. 
Moreover, 
for many of the domains where CER solutions were initially applied, 
the goal was to produce some aggregation results (e.g., averages) from
the input streams, which, in a sense, 
already constitute statistical operations.

Our analysis has identified the following classes of methods: 
automata-based methods,
probabilistic graphical models, typically based on first-order logic,
probabilistic/stochastic Petri nets 
and approaches based on stochastic grammars (usually context-free).

\subsection{Automata-based methods}
\label{approach:automata}
Most research efforts targeting the problem of uncertainty in CER are based on
extensions of crisp engines,
the majority of which employ automata. 
In this section, 
we present these approaches.
Compared to other methods, 
those based on automata seem better-suited to CER,
since input events in CER are usually in the form of streams/sequences of events,
similarly to strings of characters recognized by (Non-) Deterministic Finite Automata.
CEs are usually expressed in a declarative way (similar to SQL) with the $\mathit{sequence}$ operator playing a central role.
These expressions are subsequently transformed into automata,
using the stream of SDEs as input.
In the probabilistic versions of automata-based methods,
it is usually the SDEs that are uncertain,
accompanied by probability values with respect to their occurrence and/or attributes,
as opposed to the CE patterns.
The goal is to use these probabilistic SDEs in order to determine the probabilities of CEs.

\subsubsection*{SASE-based approaches}

SASE \cite{wu_high_performance_2006} 
and its extension, SASE+ \cite{agrawal_efficient_2008},
which includes support for $Kleene$ closure,
is an automata-based CER engine which has frequently been amended in order to support uncertainty 
\cite{shen_probabilistic_2008,kawashima_complex_2010,zhang_recognizing_2010,wang_complex_2013}.
The focus of SASE is on recognizing sequences of events through automata.
For each CE pattern,
a automaton is created whose states correspond to event types in the sequence part of the pattern,
excluding possible negations.
As the stream of SDEs is consumed,
the automaton transitions to a next state when an event is detected that satisfies the sequence constraint.
The recognized sub-sequences are pruned afterwards,
according to the other non-sequence constraints (e.g., attribute equivalences),
but, for some of these constraints,
pruning can be performed early, while the automaton is active.   
Those SDEs that triggered state transitions are recorded in a data structure
that has the form of a directed acyclic graph,
called Active Instance Stack,
allowing for quick retrieval of those sub-sequences that satisfy the defined pattern. 
SASE+ \cite{agrawal_efficient_2008} deviates somewhat from this scheme in that it employs $\mathit{NFA}^{b}$ automata,
i.e., non-deterministic finite automata with a buffer for storing matches.
For the $skip-till-any-match$ selection strategy,
where all possible SDE combinations that match the pattern are to be detected, 
the automaton is cloned when a SDE allows for a non-deterministic action.
For example, a SDE whose type satisfies a $Kleene$ operator,
may be selected, 
in which case a new automaton is created.  

Assume that our SDEs consist of events indicating whether a certain
player holds the ball, is running, dribbling, shooting or jumping.
Additionally, assume that we also have SDEs about whether the ball is in the net.
Now consider a pattern detecting an assist from a player $X$ to a player $Y$. 
This pattern could have the following simplistic definition:
\begin{equation}
\begin{aligned}
\label{eq:assist}
\mathit{assist}(X,Y,T_{3}) ::= & \mathit{hasBall}(X,T_{1});\\
& \mathit{hasBall}(Y,T_{2});\\
& \mathit{shooting}(Y,T_{3});\\
& \mathit{ballInNet}(T_{4}) 
\end{aligned}
\end{equation}
where player $X$ first has the ball, then player $Y$, who subsequently attempts a shot
and finally it is detected that the ball is in the net.
Note that, for convenience, we omit explicitly writing some production and selection predicates,
like $X{\neq}Y$.
We assume that different symbols to different variables (like $X$ and $Y$). 
We also note that this is a simplistic definition,
since the pattern does not exclude cases where there might be intervening $\mathit{hasBall}$ SDEs
between the two detected $\mathit{hasBall}$ SDEs.
A more refined pattern would have to make sure that the two detected $\mathit{hasBall}$ SDEs are consecutive,
but we use this simplistic definition for demonstration purposes,
in order to show the basic functionality of automata.

\begin{figure}[t]
\centering
{\includegraphics[scale=1.4]{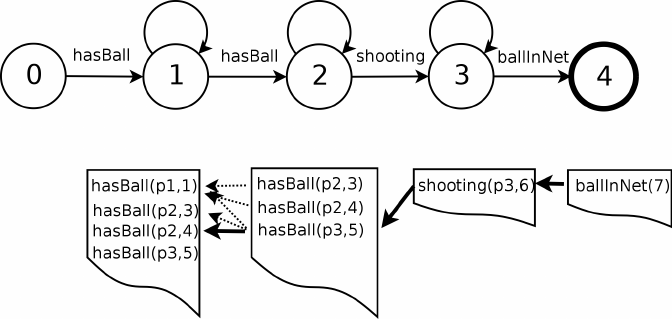}}
\caption{
NFA and Active Instance Stack as constructed by SASE for pattern \eqref{eq:assist} and example stream of Table \ref{table:assist}.
Bold arrows show a path for a full match and dashed arrows paths for partial matches.
} 
\label{fig:assist_nfa}
\end{figure}

Suppose we have a stream of probabilistic SDEs,
as the one shown in Table \ref{table:assist}.
The first column corresponds to timestamps in seconds,
while the second and third columns show SDEs and CEs respectively.
Each SDE has a probability prefix and its arguments correspond to players
(here simply denoted as $pN$) and timestamps,
except for the $\mathit{ballInNet}$ SDE,
whose only argument is the timestamp.
The $\mathit{assist}$ CEs have three arguments, 
two for the players involved and one for the timestamp.
Ignoring for the moment the probabilities,
the crisp version of SASE, 
for the $\mathit{assist}$ pattern \eqref{eq:assist},
would construct an automaton and Active Instance Stacks,
as shown in Figure \ref{fig:assist_nfa}.
Since pattern \eqref{eq:assist} is a simple sequence pattern, 
without iteration,
the NFA simply proceeds to a next state when an event of the appropriate type arrives.
The other alternative is to ignore an incoming event (self-loops)
and wait for another event in the future to satisfy the pattern.
By following the arrows in the Active Instance Stack,
the events satisfying the pattern can be efficiently retrieved.
SASE would recognize the sequence of checkmarked SDEs in Table \ref{table:assist},
producing the $\mathit{assist}(p2,p3,6)$ CE. 

\begin{table}[h]
{
\begin{tabular}{l|l|l}
\hline\noalign{\smallskip}
\textbf{Time} & \textbf{Input} & \textbf{Output} \\
\noalign{\smallskip}
\hline
$1$	& $0.9::hasBall(p1,1)$ & \\
$1$	& $0.8::dribbling(p1,1)$ & \\
$1$	& $0.95::running(p2,1)$ & \\
$3$	& $0.8::hasBall(p2,3)$ & \\
$4$	& $0.7::hasBall(p2,4)$ {\checkmark} & \\
$4$	& $0.7::dribbling(p2,4)$ & \\
$5$	& $0.9::hasBall(p3,5)$ {\checkmark} & \\
$6$	& $0.85::shooting(p3,6)$ {\checkmark} & \\
$6$	& $0.95::jumping(p6,6)$ & \\
$7$	& $0.9::ballInNet(7)$ {\checkmark} &  $P_{1}::assist(p2,p3,6)$\\
	&                     & $P_{2}::assist(p2,p3,6)$\\
	&                     & $P_{3}::assist(p1,p3,6)$\\
\ldots	& & \\
\hline
\end{tabular}
}
\caption{A stream of probabilistic SDEs from the basketball example}
\label{table:assist}
\end{table}
Note that pattern \eqref{eq:assist},
as it stands,
would also recognize two other CEs,
namely those including the $0.8::\mathit{hasBall}(p2,3)$ and $0.9::\mathit{hasBall(p1,1)}$.
Here we focus on the CE produced by the checkmarked SDEs,
as shown by the thick arrows in Figure \ref{fig:assist_nfa}.

For the probabilistic versions of SASE,
the issue is how to correctly and efficiently calculate the probability of the produced CEs.
In all of these versions 
\cite{shen_probabilistic_2008,kawashima_complex_2010,zhang_recognizing_2010,wang_complex_2013},
this probability is calculated by conceptualizing a probabilistic stream as event histories, 
produced by making a choice among the alternatives of each SDE.
In our example, 
there are two alternatives for each of the 10 SDEs in the example data stream
-- occurrence and non-occurrence --,
hence 1024 event histories. 
In \cite{kawashima_complex_2010,wang_complex_2013},
SDEs are treated as having only these two alternatives.
However, in other works (e.g., \cite{shen_probabilistic_2008}),
a SDE may have more alternatives,
corresponding to different values for the arguments of the SDE.
In \cite{zhang_recognizing_2010},
uncertainty about SDEs concerns their timestamps,
which are described by a distribution,
an issue not addressed in other works. 

The probability of a CE could be calculated by enumerating all the histories,
selecting those which satisfy the CE pattern and summing their probabilities.
The probability of a history depends on the independence assumptions that
each approach makes with respect to SDEs.
Moreover, since a full enumeration is highly inefficient,
optimization techniques are employed in order to calculate CE probabilities.

In the simplest case,
all SDEs are assumed to be independent.
In the work of \cite{kawashima_complex_2010},
where this assumption is made,
a matching tree is gradually constructed with SDEs that trigger state transitions.
By traversing the tree,
the sequence of SDEs producing a CE and its probability can be retrieved in 
a straightforward manner,
through multiplications and summation,
since all SDEs are independent. 
For our example,
the probability of the $\mathit{assist}(p2,p3,6)$ CE would be 
$0.7 * 0.9 * 0.85 * 0.9$, 
i.e. the product of the SDE probabilities. 
In this approach,
as more SDEs arrive,
probabilities can only become smaller and,
by defining a confidence threshold, 
certain branches of the tree may be early pruned.   

In \cite{shen_probabilistic_2008},
SDEs are again assumed to be independent,
but a full enumeration is avoided by using a modified version of the Active Instance Stack,
used in crisp SASE.
The concept of lineage, borrowed from the field of probabilistic databases \cite{benjelloun2006uldbs},
is used for calculating CE probabilities. 
A similar approach is used by Wang et al. \citeyear{wang_complex_2013},
where the assumption of complete SDE independence is relaxed and some SDEs
may follow a first-order Markov process.
In this case,
the edges of the Active Instance Stack are annotated with the conditional probabilities.
If we assume that 
\begin{equation*}
\begin{aligned}
P(\mathit{ballInNet}(7)|\mathit{shooting(p3,6)})=0.95
\end{aligned}
\end{equation*}
then 
\begin{equation*}
\begin{aligned}
P(\mathit{ballInNet}(7),\mathit{shooting(p3,6)}) = & P(\mathit{shooting(p3,6)})*\\
& P(\mathit{ballInNet}(7)|\mathit{shooting(p3,6)}) = \\
& 0.85*0.95
\end{aligned}
\end{equation*}
hence 
\begin{equation*}
\begin{aligned}
P(\mathit{assist}(p2,p3,6)) = 0.7 * 0.9 * 0.85 * 0.95
\end{aligned}
\end{equation*}

Note that the conditional probability tables in this approach are based on event types,
e.g., $P(\mathit{ballInNet}|\mathit{shooting})$ for all $\mathit{ballInNet}$ and $\mathit{shooting}$.
For every specific SDE dependent on another SDE, probabilities must be explicitly provided. 
In this work,
hierarchies of CEs are also allowed.

In \cite{zhang_recognizing_2010}, the issue of imprecise timestamps is
addressed, while all the other attributes have crisp values.
Again, SDEs are assumed to be independent. 
Complete enumeration of all possible worlds is avoided by employing an incremental,
three-pass algorithm through the events in order to construct event matches and
their intervals.
This work was later extended \cite{zhang_complexity_2014}, by adding $negation$ and
$Kleene\ plus$ and by allowing for user-defined predicates.

All of these SASE-based methods perform marginal inference 
and use confidence thresholds for pruning results that fall below them.
Only one attempts to increase performance through distribution.
To the best of our knowledge, 
the work of \cite{wang_complex_2013} is one of the very few developing a CER system which is both
probabilistic and distributed
(PADUA, described later, is the only other such method).

\subsubsection*{Other automata-based approaches}

Non SASE-based approaches have also appeared.
A recognition method that models activities
using a stochastic automaton language is presented by Albanese et al. \citeyear{AlbaneseMPSU07}.
In this case,
it is not the SDEs that are probabilistic,
but the state transitions of the automaton,
similarly to Markov chains. 
A possible world is now essentially defined over activity occurrences that are targeted for recognition, 
i.e. CEs.
This method was later extended \cite{AlbaneseMPPS11}, 
in order to identify situations that cannot be satisfactorily explained by any
of the known CEs.
In particular, 
the stochastic automaton model is extended with temporal constraints, 
where subsequent SDEs can occur within a user-defined temporal interval.
Using possible-worlds based modeling, 
the method finds (partially) unexplained activities.
This is the only automata-based method that can perform both marginal and MAP inference.

Another extension of \cite{AlbaneseMPSU07}
is the PADUA system \cite{MolinaroMPPRS14}. 
PADUA employs Probabilistic Penalty Graphs  
and extends the stochastic automaton presented in \cite{AlbaneseMPSU07} with noise degradation.
The edges that connect subsequent events in a Penalty Graph, forming the structure of a
CE, are associated with a probability (noise) value that degrades the belief of
the CE when other events intervene.
As a result, under such situations, the CE is being recognized, but with
reduced probability.
Besides CER, the method can find patterns of events that do
not belong to the set of known CEs.
Additionally, for purposes of scalability, Probabilistic Penalty Graphs
can be combined by merging common sub-Graphs, indexing and executing them in parallel.

Lahar \cite{re_event_2008} constitutes one of the earliest proposals
and is based on the Cayuga engine \cite{demers_towards_2006}.
Events are modeled by first-order Markov processes. 
The supported queries are categorized in three different classes of increasing complexity.
For the first two types of queries (Regular and Extended Regular),
automata are used for recognition.
For the most complex queries (Safe),
in which variables are not shared among all of the conditions
(e.g., as in pattern \eqref{eq:assist}, with $X$ and $Y$),
a version of the Probabilistic Relational Algebra \cite{fuhr_probabilistic_1997} is used.
A method which attempts to overcome the strict markovian hypothesis of Lahar 
and apply certain optimizations, such as early pruning, may be found in \cite{chuanfei_complex_2010}.

\subsubsection*{Commentary}
\label{comments_auto}

Automata-based methods focus on recognizing sequences of events,
in which some of those events may be related, 
via their attributes, 
to other events of the sequence.
In general, 
time representation is implicit.
As a result,
and with the exception of \cite{AlbaneseMPPS11},
they do not include explicit temporal constraints,
such as concurrency or inequalities between timestamps,
e.g., a constraint like $T_{4} - T_{1} \le 24\ seconds$ in pattern \eqref{eq:assist}
to make sure than the recognized activity takes place within a single offense.
$Windowing$ is the only temporal constraint allowed.
Moreover, 
they only address the issue of data uncertainty
(the exception here is again \cite{AlbaneseMPPS11}),
lacking a treatment of other types of uncertainty, such as pattern uncertainty,
and model relatively simple probabilistic dependencies between events.

To illustrate a case where concurrency and more complex dependencies may be required,
consider a pattern trying to detect an attempted block by a defender from the stream of Table \ref{table:assist}:
\begin{equation}
\begin{aligned}
\label{eq:attempted_block}
\mathit{attempt\_block}(Y,T) ::= & \sigma_{\mathit{opponents}(X,Y)}( \\
& \mathit{shooting}(X,T) \ \wedge \\
& \mathit{jumping}(Y,T) \ \wedge \\
& \mathit{close}(X,Y,T) \ )
\end{aligned}
\end{equation}
where player $X$ is shooting at the same time that player $Y$ is jumping,
the distance between them is small at that time
(we assume image recognition can provide such information as $\mathit{close}$ SDEs)
and the two players belong to different teams, i.e., they are $\mathit{opponents}$.
Such a pattern would require explicit temporal constraints or, at least,
an implicit constraint about concurrent events,
a feature generally missing in automata-based methods.
Moreover, 
$\mathit{jumping}$ is clearly dependent on $\mathit{shooting}$
(a player usually jumps at the same time or after another player shoots
but not while all other players run),
yet this dependence cannot be captured by assuming a Markov process that generated those events.
Note also that the above pattern makes use of the $\mathit{opponents}$ predicate,
assuming that the engine can take into account such background knowledge that is not part of the SDE stream.
Such knowledge is relatively easy to model in logic-based systems,
but the automata-based ones presented above have no such mechanism.

Now assume we want to express a rule stating that if two players are $\mathit{close}$ to
each other at the current timepoint, then they are likely to be close at the
next timepoint (a first-order Markov assumption).
This is not an event we would like to detect in itself,
but domain knowledge which we would like our system to take into account.
Such rules may be helpful in situations where SDEs may suddenly be missing,
for example due to some sensor failure,
but the activity has not ceased. 
We could introduce the following two patterns:
\begin{align}
& \label{eq:close_1}
\begin{mysplit}
1::\mathit{close}\_m(X,Y,T) ::=  \mathit{close}(X,Y,T)
\end{mysplit}\\
& \label{eq:close_2}
\begin{mysplit}
0.6::\mathit{close\_m}(X,Y,T) ::=  \sigma_{\mathit{next}(T,T_{previous})}( \\
\qquad\qquad\qquad\qquad\qquad\qquad \mathit{close\_m}(X,Y,T_{previous})\ )
\end{mysplit}
\end{align}
and use the $\mathit{close\_m}$ predicate instead of $\mathit{close}$ in the
definition of pattern \eqref{eq:attempted_block} for $\mathit{attempt\_block}$.
The first of these patterns simply transfers the ``detection'' probability of
$\mathit{close}$ to that of $\mathit{close\_m}$ (pattern probability is $1$), 
whereas the second one expresses the Markov assumption.
In automata-based methods where Markov assumptions are allowed,
the conditional probabilities need to be provided for every ``ground'' pair of SDEs.
Uncertain patterns allow us to describe such dependencies in a more succinct manner,
as ``templates''.

Assume also that we need some patterns to detect maneuvers in which the offender
attempts to avoid the defender.
Two of these patterns could be the following:
\begin{align}
& \label{eq:surpass_1}
\begin{mysplit}
0.9::\mathit{avoid}(X,Y,T_{2}) ::=  \mathit{waiting}(X,Y,T_{1});  \\
\qquad\qquad\qquad\qquad\qquad\quad \mathit{crossover\_dribble}(Y,T_{2})
\end{mysplit}\\
& \label{eq:surpass_2}
\begin{mysplit}
0.7::\mathit{avoid}(X,Y,T_{2}) ::=  \mathit{waiting}(X,Y,T_{1}); \\
\qquad\qquad\qquad\qquad\qquad\quad \mathit{running}(Y,T_{2})
\end{mysplit}
\end{align}
where $\mathit{crossover\_dribble}$ and $\mathit{waiting}$ are assumed to be CEs detected by their
respective patterns.

In this case,
we have a hierarchy of CEs,
defined by probabilistic patterns,
starting with the SDEs,
on top of them the $\mathit{waiting}$ and $\mathit{crossover\_dribble}$ CEs
and finally the $\mathit{avoid}$ CE.
An efficient mechanism for propagating probabilities among the levels of the CE hierarchy would be required.
Among the presented methods,
CE hierarchies are allowed only in \cite{wang_complex_2013}.
Moreover,
combining rules would also be required,
both for patterns \eqref{eq:close_1} -- \eqref{eq:close_2} and patterns \eqref{eq:surpass_1} -- \eqref{eq:surpass_2},
i.e. functions for computing the probabilities of CEs with multiple patterns
(common head, different bodies).
For example, pattern \eqref{eq:surpass_1} provides the probability of $\mathit{avoid}$ given
$\mathit{waiting}$ and $\mathit{crossover\_dribble}$ and pattern \eqref{eq:surpass_2} the probability of $\mathit{avoid}$ 
given  $\mathit{waiting}$ and $\mathit{running}$,
but we do not know this probability given all of the lower-level CEs. 
A combining rule could help us in computing such probabilities,  
without adding them explicitly. 

\subsection{First-order logic \& Probabilistic Graphical Models}
\label{approach:pgms}

Another line of research revolves around methods which employ 
probabilistic graphical models (PGMs) in order to handle uncertainty.
These models take the form of networks whose nodes represent random variables 
and edges encode probabilistic dependencies.
The two main classes of PGMs used in CER are Markov Networks and Bayesian Networks,
the former being undirected models whereas the latter are directed.
When used for CER,
Markov Networks may be combined with first-order logic,
in which case they are called Markov Logic Networks (MLNs).
The nodes in a MLN represent ground logical predicates,
as determined by the (weighted) formulas that express CE patterns.
When Bayesian Networks are used,
the nodes usually represent events (SDEs and CEs).

\subsubsection*{Markov Logic Networks}

Since their first appearance \cite{richardson_markov_2006}, 
Markov Logic Networks (MLNs) have attracted increasing attention as a tool 
that can perform CE recognition under uncertainty.
MLNs are undirected probabilistic graphical models which encode
a set of weighted first-order logic formulas
(for a comprehensive description of MLNs, see \cite{domingos_markov_2009}).
CEs are expressed as logic formulas which may even contain existential and universal quantifiers
(although the former can prove quite expensive), 
as with first-order logic.
However, in first-order logic,
a possible world 
(i.e., an assignment of truth values to all ground predicates)
that violates even one formula is considered as having zero probability.  
In order to avoid this strict requirement,
the methods described in this section may allow a formula to be ``soft'',
meaning that it is accompanied by a probability/weight value,
indicating how ``strong'' we need it to be compared to other formulas.
As a consequence,
possible worlds can have non-zero probability,
even when violating some formulas,
albeit a lower one than those without violations.

\begin{figure}[t]
\centering
{\includegraphics[scale=1.5]{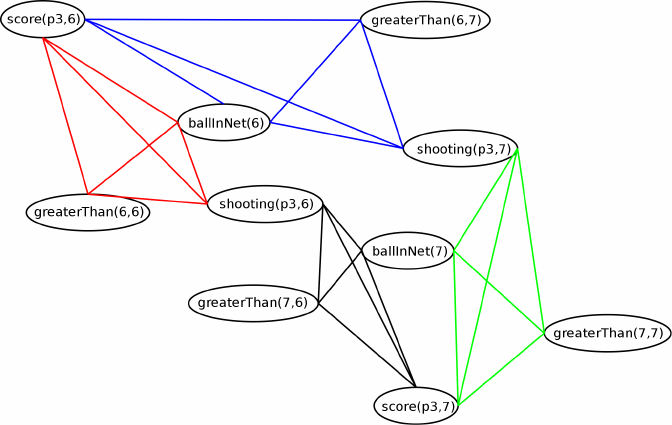}}
\caption{Ground Markov Logic Network constructed for rule \eqref{eq:score_fol} and for part of the stream in Table \ref{table:assist}.} 
\label{fig:score_mln}
\end{figure}

For a simple example of how a formula in first-order logic is encoded as a MLN,
consider a formula in first-order logic about a player $\mathit{scoring}$:
\begin{equation}
\begin{aligned}
\label{eq:score_fol}
\mathit{score}(X,T_{2}) & \leftarrow & \\
& \forall X,T_{1},T_{2}  & \\
 & \mathit{shooting}(X,T_{1}) \ \wedge & \\
 & \mathit{ballInNet}(T_{2}) \ \wedge & \\
 & \mathit{greaterThan}(T_{2},T_{1}) \  & \\
 \end{aligned}
\end{equation}
Note that in first-order logic,
it is not possible to directly perform numerical calculations and time
(in-)equalities must be explicitly provided in the form of predicates,
as above, with $\mathit{greaterThan}$.
If we take into account only one player ($p3$) and only two timepoints (6,7)
from the stream of Table \ref{table:assist},
then the MLN corresponding to this formula 
would be the one shown in Figure \ref{fig:score_mln}.
Each possible ground predicate is represented as a node
and edges exist between nodes that appear together in a ground formula.
For example, one possible grounding of the above formula, for $X=p_{3}$, $T_{1}=6$ and $T_{2}=7$, is the following:
\begin{equation}
\begin{aligned}
\label{eq:score_fol_1}
\mathit{score}(p_{3},7) & \leftarrow & \\
 & \mathit{shooting}(p_{3},6) \ \wedge & \\
 & \mathit{ballInNet}(7) \ \wedge & \\
 & \mathit{greaterThan}(7,6) \  & \\
 \end{aligned}
\end{equation}
which corresponds to the bottom left clique of nodes in Figure \ref{fig:score_mln} (black edges).
The other three cliques (red, blue, green) correspond to the remaining possible groundings of the formula, 
i.e., the remaining combinations of the two involved timepoints ($T_{1}=6$/$T_{2}=6$, $T_{1}=7$/$T_{2}=6$ and $T_{1}=7$/$T_{2}=7$ respectively).
Note that this is a direct and naive way of obtaining a ground MLN.
Clever algorithms can prune unnecessary parts of the graph.

There is a substantial body of work on CER with MLNs 
\cite{BiswasTF07,HelaouiNS11,TranD08,morariu11,sadilek2012,skarlatidis_probabilistic_2011,skarlatidis_probabilistic_2015,SongSRAI13,SongICMI13,kanaujia_complex_2014}.
All of these methods are concerned with human activity recognition, 
with input events derived mostly from video sources (less frequently from
GPS or RFID traces).
As a result,
many of them have developed solutions that are domain dependent. 
Here we focus on those representative papers that are more closely related to CER,
by providing a more generic way for handling events.
For example, in \cite{morariu11} and \cite{SongSRAI13}, 
where Allen's Interval Algebra \cite{allen_maintaining_1983} is used
and in \cite{skarlatidis_probabilistic_2011,skarlatidis_probabilistic_2015},
where a version of the Event Calculus \cite{kowalski_logicbased_1986} is used.
With the use of such formalisms,
temporal constraints are not captured in the simplistic way implied by the 
$\mathit{greaterThan}$ predicate.
Instead,
built-in predicates about the temporal relations of events are provided,
e.g., the $\mathit{after}$ relation in Allen's Interval Algebra,
for indicating that an event succeeds in time another event.

Contrary to automata-based solutions,
MLNs focus on encoding probabilistic rules.
This allows both for incorporating background knowledge and for building
hierarchies of CEs with correct probability propagation.
On the other hand, they use the less intuitive weights instead of probabilities,
which indicate how strong a rule is compared to the others.
While it might be possible for certain simple domains to manually define weights,
usually a learning phase is required to optimize them.

As far as data uncertainty is concerned, 
it is possible to include probabilistic SDEs as well.
Since a node in a MLN is not directly associated with a probability
(it is the formulas/graph cliques that have weights),
these SDEs must be expressed as formulas too.
Such formulas connect each observed SDE with a ``generated'' SDE,
with an appropriate weight 
(see the \textit{Commentary} section below for more details).
Moreover, MLNs allow for more complex reasoning about SDEs.
For example,  
in the work of \cite{TranD08},
besides handling noisy SDEs,
missing SDEs may be inferred through rules about what must have happened for an event to have occurred.

A similar approach is proposed by Morariu and Davis \citeyear{morariu11},
where the Interval Algebra is employed and the most consistent sequence of CEs are determined, 
based on the observations of low-level classifiers.
In order to avoid the combinatorial explosion of possible intervals, 
as well as to eliminate the existential quantifiers in CE definitions, 
a bottom-up process eliminates the unlikely event hypotheses. 
The elimination process is guided by the observations and the interval
relations of the CE patterns.

In \cite{SongSRAI13,SongICMI13}, MLNs are again combined with Allen's Interval Algebra,
upon which a set of domain-independent axioms is proposed. 
Abstraction axioms define hierarchies of events in which an instance of an event with a given type
is also an instance of all abstractions of this type.
Prediction axioms express that the occurrence of an event implies the
occurrence of its parts.
Constraint axioms ensure the integrity of the (temporal) relations among CE and
its parts.
Finally, abduction axioms allow CE to be inferred on the basis
of their parts, 
i.e. inferring when some events are missing.
In this case, SDEs are not probabilistic. 
 
\begin{figure}[t]
\centering
{\includegraphics[width=130mm]{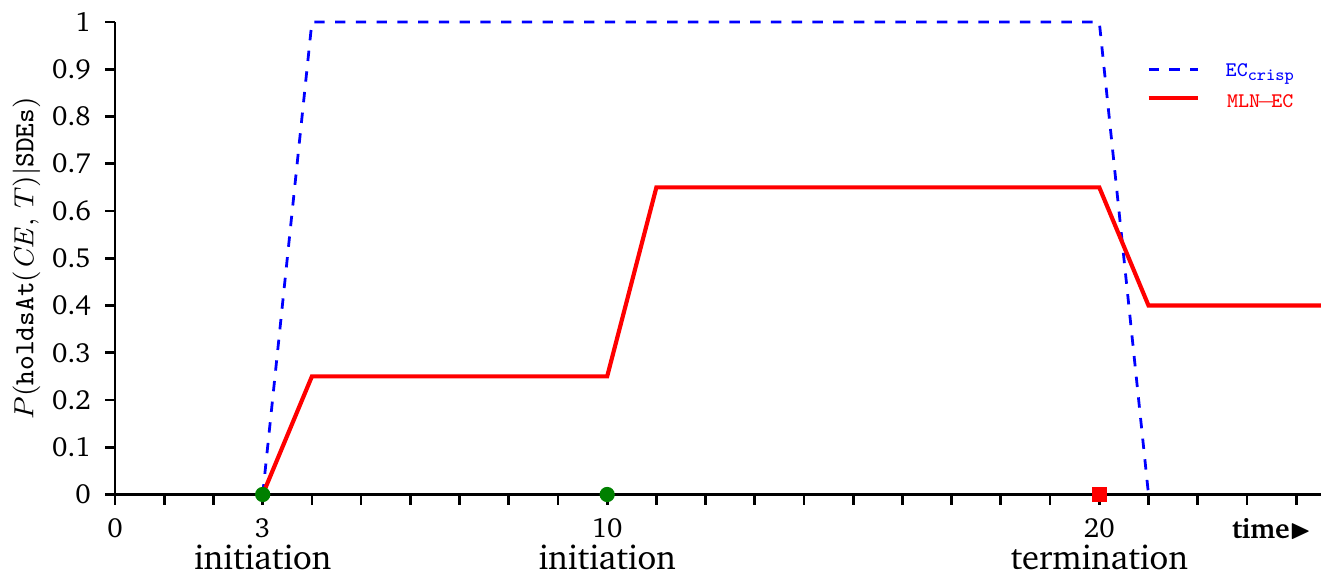}}
\caption{Probability increasing for every initiation point in MLN-EC 
$Y$ axis shows the probability that a CE is true.
The dashed blue line corresponds to a crisp version of the Event Calculus.
The solid red line corresponds to a probabilistic version with MLNs.} 
\label{fig:mln_ec}
\end{figure}

The work of \cite{skarlatidis_probabilistic_2011,skarlatidis_probabilistic_2015} 
represents one of the first
attempts to provide a general probabilistic framework for CER via MLNs (MLN-EC). 
In order to establish such a framework,
a version of the Event Calculus is used whose axioms are domain-independent.
Combined with the probabilistic domain-dependent rules,
inference can be performed regarding the time intervals during which activities of
interest (fluents in the terminology of the Event Calculus) hold.
The problem of combinatorial explosion due to the multiple time-points that need
to be taken into account is addressed by employing a discrete version of the Event Calculus,
using only integer time-points and axioms that relate only successive time-points.
For similar reasons, existential quantifiers are not allowed. 
Due to the law of inertia of the Event Calculus 
(something continues to hold unless explicitly terminated or initiated with a different value),
this model increases the probability of an inferred event every time its
corresponding rule is satisfied and decreases this probability whenever its
terminating conditions are satisfied,
as shown in Figure \ref{fig:mln_ec}. 
In \cite{skarlatidis_probabilistic_2013} the Event Calculus was again used, 
but this time the focus was on probabilistic SDEs rather than probabilistic rules. 
The ProbLog framework was used \cite{kimmig_implementation_2011},
a probabilistic extension of the logic programming language Prolog.
ProbLog allows for assigning probabilities to SDEs, i.e. for probabilistic facts. 
Intuitively, the success probability of a query (CE rule) is the probability that this query is provable,
starting from the (probabilistic) facts and any other rules that might be present as background knowledge.
The axioms of the Event Calculus were expressed in ProbLog as background knowledge
and a set of CE rules for human activity recognition from video sources was tested.
The method exhibited improved accuracy performance with respect to crisp
versions of the Event Calculus, 
when tested against noisy SDE streams.

The work of \cite{brendel2011PEL,selman2011PEL} presents the Probabilistic Event Logic 
that defines a log-linear model from a set of weighted formulas.
It does not directly employ MLNs but it is very close in spirit.
The formulas that describe CEs are represented in Event Logic,
a formalism for defining interval-based events
\cite{siskind_grounding_2001}, by employing the operators of Allen's Interval Algebra.
Each formula defines a soft constraint over some events, 
in a manner similar to MLNs.
However, instead of building a ground network with all the time variables,
the inference algorithm works with a specialized data structure, called \textit{spanning intervals}.
This data structure allows for a compact representation of event occurrences that satisfy a formula.
The inference algorithm can work with set operations on intervals and performs local-search,
based on MaxWalkSAT \cite{kautz1997MaxWalkSAT}.
Moreover, a method is presented for learning the weights of the formulas.

A probabilistic activity language
for expressing CEs on top of an image processing suite is proposed by Albanese et al. \citeyear{AlbanesePADS10}.
This work does not employ graphical models,
but we include it here because it is based on first-order logic.
CEs are defined by users in first-order logic.
These definitions are flexible enough to incorporate both universal and existential quantifiers, 
as well as alternations of quantifiers.
As far as the input SDEs are concerned,
they can be either Boolean (with the usual true/false values) 
or probabilistic, accompanied by an occurrence probability.
The method can handle both instantaneous events and events that span over intervals.
In order to compute the probabilities of CEs,
the dependencies between the SDEs must be modeled as well.
For this purpose,
triangular norms \cite{Fagin96} are used.
Triangular norms are binary functions
that can model probabilistic dependencies more general than those of exclusivity or independence
For example, the \emph{minimum} function can be used as such a norm 
($\mathit{min(x,y)}$, where $x$ and $y$ are the probabilities of the two SDEs whose dependence we want to model).
The off-line detection algorithm works by recursively decomposing a CE formula into its parts, 
finding the valid substitutions and keeping those that have a probability above a certain threshold.
An on-line algorithm is also presented,
which can also detect partially completed activities.

\subsubsection*{Bayesian Networks}
Bayesian Networks are directed graphical models (in contrast to MLNs, which are undirected) 
whose structure can encode probabilistic dependencies between random variables,
represented as nodes in the network.
When used for CER,
the nodes of the network usually correspond to SDEs and/or CEs.
Therefore, the network encodes how SDEs and CEs depend on each other.
In Bayesian Networks, these dependencies are more ``localized'',
in the sense that computing the probability of a node does not require 
some kind of global knowledge of the network,
as with the partition function of MLNs.
The rest of this section describes CER methods whose probabilistic model
is encoded via Bayesian Networks.

The work presented in
\cite{wasserkrug2008complex,wasserkrug_efficient_2012,wasserkrug_model_2012}
employs the technique of knowledge-based model construction
(KBMC), whereby knowledge representation is separated from the inference 
process.
Each event is assigned a probability, denoting how probable it is that the
event occurred with specific values for its attributes.
In turn, CE patterns are encoded in two levels, 
with a selection operation performing an initial
filtering, mostly based on event type, 
followed by a pattern-detection schema for more complex operations, 
based on temporal and attribute constraints.
The selection mechanism imposes certain independence properties on the network.
CEs are conditioned only on selectable lower-level events
(as determined by the selection operation), 
preventing the network from being cluttered with many dependency edges.
This framework is not limited to representing only propositional or even first-order 
knowledge.
It could potentially handle higher-order knowledge,
since the pattern-matching step may, in principle, 
be defined in any kind of language.
However, the system presented \cite{wasserkrug2008complex,wasserkrug_efficient_2012,wasserkrug_model_2012}
allows only predicates expressing temporal constraints
on event timestamps and equality constraints on event attributes.
 
The calculation of probabilities for the CEs is done by 
a Bayesian network that is dynamically constructed upon each new event arrival.
The nodes of the network correspond to SDEs and CEs.
First, SDEs are added.  
Nodes for CEs are inserted only when a rule defining the CE is crisply satisfied, 
having as parents the events that triggered the rule,
which might be SDEs or even other CEs, in case of hierarchical CE patterns.
The attribute values of the inferred CEs are determined 
by mapping expressions associated with the corresponding rule,
i.e. functions mapping attributes of the triggering events to attributes of the inferred event.  
In order to avoid the cost of exact inference, 
a form of sampling is followed, 
that bypasses the construction of the 
network by sampling directly according to the CE patterns.

A more recent effort extends the TESLA \cite{cugola_tesla_2010} event 
specification language with probabilistic modeling, 
in order to handle the uncertainty both in input SDEs and in the
CE patterns \cite{Cugola2014CEP2U}.
The semantics of the TESLA language is formally specified by using a first-order 
language with temporal constraints that express the length
of time intervals.
At the input level, 
the system, called CEP2U, supports uncertainty regarding the occurrence of the
SDEs, as well as the uncertainty regarding the content of the SDEs.
SDEs are associated with probabilities that indicate a degree of confidence,
while the attributes of a SDE are modeled as random variables
with some measurement error.
The probability distribution function of the measurement error is assumed to be known.
The method also models the uncertainty of CE patterns, 
by automatically building a Bayesian network for each rule.
The probabilistic parameters of the network are manually estimated by domain
experts.

Other methods based on Bayesian networks, 
which could be used for CE recognition include Bayesian logic programming \cite{getoor_introduction_2007}, 
relational Bayesian Networks \cite{jaeger_relational_1997} 
and relational dynamic Bayesian Networks 
\cite{sanghai_relational_2005}.
Towards this direction, 
Dynamic Bayesian Networks have been extended using first-order logic
\cite{manfredotti2009RBN,manfredotti2010RBN}.
A tree structure is used, where each node corresponds to a first-order logic
expression, e.g., a predicate representing a CE, 
and can be related to nodes of the same or previous time instances.
Compared to their propositional counterparts, 
the extended Dynamic Bayesian Networks methods can compactly represent CE that involve various
entities.

\subsubsection*{Commentary}
\label{comments_pgms}

PGMs,
such as Markov Logic Networks and Bayesian Networks,
can provide a substantial degree of flexibility 
with respect to the probability distributions that they can encode.
On the one hand,
they are very expressive and they don't require restrictive independence assumptions to be made.
On the other hand,
this increased flexibility comes at a cost with respect to efficiency.
In general,
a rule which references certain random variables implies that, 
before inference can begin, the cartesian product of all the values of these
variables needs to be taken into account.
For human activity recognition, 
one may assume that the number of persons involved in a
scene is relatively limited.
However, this is not the case for all domains.
For a fraud detection scenario, involving transactions with credit cards,
a CER system may receive thousands of transactions per second, 
most of them having different card IDs.
The demands of CER exacerbate this
problem, since time is a crucial component in these cases.
The possible combinations of time points with the other random variables
can quickly lead to intractable models.
All of the papers discussed in this section employ low-arity (or
even 0-arity) predicates, 
whose arguments have small domain sizes, except for that of time. 
In order to reduce the unavoidable complexity introduced by the existence of time, 
they develop special techniques, 
such as the bottom-up technique in \cite{morariu11}.

With respect to probabilistic SDEs,
although they can be incorporated into graphical models,
correctly encoding their dependencies can be far from obvious,
especially with MLNs.
Assume we want to assign an occurrence probability of $80\%$ to the $\mathit{close}$ SDE 
of rules \eqref{eq:close_1}--\eqref{eq:close_2}.
With MLNs, we could replace rule \eqref{eq:close_1} with an equivalence rule, such as:
\begin{equation}
\begin{split}
\label{eq:close_3}
1.39 \ \forall X,Y,T \ \mathit{close}(X,Y,T) \leftrightarrow \mathit{close\_m}(X,Y,T)
\end{split}
\end{equation}
with the appropriate weight (log-odds of occurrence and non-occurrence
probabilities), where $\mathit{close}$ are the observed SDEs and $\mathit{close\_m}$ the
inferred probabilistic events to be used in other rules, such as \eqref{eq:attempted_block}.
It would not be sufficient to directly express rules \eqref{eq:close_2} and \eqref{eq:close_3} 
in first-order logic and use them to construct an MLN, 
since the dependencies introduced would mean that the marginal probability of the 
$\mathit{close}$ SDE could be affected by the probability of the 
$\mathit{close\_m}(X,Y,T_{previous})$ predicate.
Bayesian Networks could be used to avoid such problems,
due to their directionality.
On the other hand,
MLNs can be trained as discriminative models,
which means that it is not necessary to explicitly encode all the probabilistic dependencies.

\subsection{Petri Nets}
\label{approach:petri}
\begin{figure}[t]
\centering
{\includegraphics[scale=1.8]{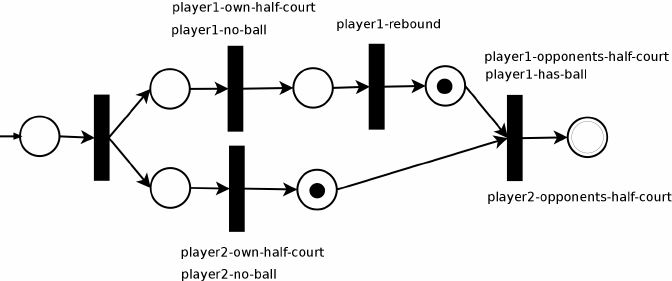}}
\caption{Basketball transition pattern as a Petri Net} 
\label{fig:petri}
\end{figure}

In order to address issues of concurrency and synchronization, 
some methods have employed probabilistic extensions of Petri Nets.
Formally, a Petri Net may be described as a bipartite directed graph (see \cite{murata_petri_1989,peterson_petri_1981} for more complete discussions).
It has two types of nodes:
\begin{itemize}
 \item \emph{ Place nodes}, representing states of the modeled activity (usually depicted as circles). Each such node may hold multiple \emph{tokens}, usually depicted as small, black circles inside place nodes.
 \item \emph{Transition nodes} (usually depicted as rectangles), which, as the name suggests, connect place nodes. 
 A transition node is said to be enabled when all of its input place nodes have tokens, 
 in which case the node may fire.
 Conditions may be imposed on transition nodes to determine exactly when they should fire. Firing removes all enabling tokens from the input place nodes
 and writes a new token at each output node.
\end{itemize}
A \emph{marking} is a function that assigns tokens to place nodes.  
When a transition fires,
the marking of the Petri Net changes to a new one.
A simple example of a rule describing a transition in basketball from defense to offense, encoded as a Petri Net,
is shown in Figure \ref{fig:petri}.
The marking in this figure means that $\mathit{player1}$ has grabbed a rebound and $\mathit{player2}$
is in his team's half-court and does not have the ball. 
It is important to note that, for the activity to complete,
all three conditions of the final transition node must hold, 
i.e., $\mathit{player1}$ must have the ball and both players must be in the opponent team's half-court. 
This is a transition node that enforces synchronization between the two players' activities. 
The ability to model constraints about synchronization and concurrency is a powerful feature of Petri Nets.
 
A probabilistic extension to Petri Nets has been proposed in
\cite{AlbaneseCCMPSU08}, 
for recognizing CEs that represent human activities. 
A Petri Net expresses a CE and is formed by SDEs  that are connected with constraints, like temporal durations.
The SDEs and other possible constraints are encoded as conditions on transition nodes.
Special \emph{dead-end} place nodes are used in order to represent forbidden actions that must immediately trigger an alarm.
The transition from one state to another is associated with a probability value.
Note that there are might be multiple possible transitions from a certain place node
(in our example, after $\mathit{player1}$ grabs the rebound, he could also make a pass to $\mathit{player2}$ as an alternative).
The sum of the probabilities of these transitions is $1$.
Given a sequence of SDEs, the method can identify segments of the sequence in
which a CE occurs with probability above a specified threshold or infer the
most likely CE in the sequence.

A stochastic variant of Petri-Nets that models the
uncertainty of input SDEs is proposed in by Lavee et al. \citeyear{Lavee13PFPN}, 
an issue not addressed in \cite{AlbaneseCCMPSU08}.
Specifically, SDEs are recognized with some certainty, using lower level classification algorithms.
CEs are represented by Petri Nets, in terms of SDEs that are associated with certainties and temporal constraints.
The probability of a CE being recognized is computed through a Bayesian particle filter approach.
It combines a dynamic model for the probability of moving to a new state (marking) given the previous one
and a measurement model for the probability of observing SDEs, given the current state.
However, the transitions themselves cannot be assigned arbitrary probabilities by the user.
If there are paths from one marking to other markings (via the SDEs), 
then the dynamic model assigns equal probabilities to all these paths.

\subsubsection*{Commentary}
\label{comments_petri}

One limitation of the approaches that use Petri Nets is the lack of a
mechanism for modeling a domain in a truly relational manner, 
i.e. by allowing relations to be defined between attributes of events.
These methods treat events as 0-arity predicates, 
related only through temporal constraints, as implied by the structure of the Petri Net.
As is the case with automata too,
Petri Nets tend to make a significant number of independence assumptions.
The domain on which they have been tested is that of human activity recognition,
in which the sequential nature of activities allows for the adoption 
of first-order Markov models.
As far as inference is concerned,
both MAP and marginal are possible and optimization techniques,
such as confidence thresholds and approximate inference have been employed.

\subsection{Grammars}
\label{approach:grammars}
A number of research efforts have focused on syntactic approaches to CE recognition.
These approaches typically convert a stream of input SDEs to a stream of
symbols upon which certain user-defined rules may be applied.
Rules are defined via a stochastic grammar 
(see \cite{stolcke_efficient_1995} for a description of stochastic context-free grammars) 
in order to take uncertainty into account.
This can be achieved by assigning a probability value to each production of the grammar.
The probabilities of all productions with the same non-terminal symbol on their left-hand side 
must sum to $1$.
Stochastic grammars, like the ones described in the rest of this section,
have some attractive features.
For example, it is easy to define CEs within a single hierarchical grammar.
Moreover, they can provide probability evaluations even for ``partial matches'' of a CE,
e.g., for its prefixes,
a useful feature for predicting events that might follow.

A two-step approach along these lines is proposed in \cite{ivanov_recognition_2000}. 
Low-level detectors, based on Hidden Markov Models, are used to generate a symbol stream which is
then fed into a parser constructed from a stochastic context-free grammar.
This parser in turn employs stochastic production rules in order to determine
the probability of a high-level event.
Similar approaches may be found in
\cite{moore_recognizing_2002} and in \cite{minnen_expectation_2003}.
Note though that the latter adds context-sensitive symbols to the grammar,
i.e., symbols/events may have arguments.

Ryoo and Aggarwal \citeyear{RyooA09} proposed a hierarchical method
that combines a syntax for representing the CE patterns of \cite{ryoo2006} 
with probabilistic recognition.
A syntax similar to that of context-free grammars is used for describing CEs,
but the actual recognition is treated as a constraint satisfaction problem.
The method aims to probabilistically detect the time intervals in which CEs
occur.
Input SDEs are associated with probabilities, indicating a degree of belief. 
Based on a context-free grammar representation scheme, 
CE patterns are expressed in terms of other events (SDEs or CEs) and form a
hierarchical model.
Furthermore, events are related with temporal interval, logical and
spatial constraints \cite{allen_maintaining_1983}.
In situations where the input stream is incomplete, the method generates the
missing SDEs with low confidence.
The probability of a CE is calculated by exploiting the dependency information
between the CE and its sub-events, as encoded in the hierarchy tree.
This method for calculating probabilities is similar to that of Bayesian networks,
the main difference being that siblings are not assumed to be conditionally independent.
When the calculated probability of the CE is above a specified threshold, 
it is considered as recognized.

Moving up to context-sensitive grammars, 
the work in \cite{pei_learning_2013} represents CEs as temporal And-Or Graphs (T-AOG)
and attempts both to recognize events from video streams and to infer a person's intentions.
In a T-AOG, terminal nodes represent atomic actions (SDEs),
i.e., elementary sequences of spatial relations between agents and objects.
On top of these atomic actions, 
a hierarchy of events is constructed, using And-nodes and Or-nodes.
And-nodes define temporal (sequential) relations  that regulate the durations of its sub-events.
All of the sub-events of an And-node must occur.
On the contrary, 
Or-nodes represent different options between events and each such option has an associated probability
(for a complete discussion of And-Or Graphs, see \cite{zhu_stochastic_2007}).
Given a sequence of SDEs,
a parsing algorithm incrementally constructs all the possible partial graphs that can interpret the input sequence. 
In order to limit the number of partial graphs,
those whose probability falls below a certain threshold are pruned.
The probability of a graph is computed by taking into account
the probabilities of the SDEs,
the frequency that an Or-node follows a path
and the probability that an And-node's temporal relations are satisfied.
As a graph is incrementally constructed,
the system can also make predictions about the events that are expected to follow.
Finally, this work is one of the few that learns CE patterns,
by using an unsupervised learning algorithm to build the T-AOG.

\subsubsection*{Commentary}
\label{comments_grammars}

In \cite{ivanov_recognition_2000} the SDEs (terminal symbols) are 
represented as 0-arity,
hence no relations may be defined on attributes.
Since no event attributes are allowed,
it is not possible to define probabilities on attributes either.
Moreover, when defining a (production) rule,
all the possible sub-scenarios (expansions) must be explicitly stated,
with probability values that sum to $1$.
For example, a rule for detecting the $avoid$ event, as in rule \eqref{eq:surpass_1},
cannot be ``simply'' written, as:
\begin{equation}
\begin{aligned}
\label{eq:prod_ex_1}
0.9:: & Avoid\_Player1\_Player2 & \rightarrow \\ & Waiting\_Player1\_Player2, & \\
& Crossover\_Dribble\_Player2 &
\end{aligned}
\end{equation}
This rule has a probability of $0.9$ and therefore we need extra scenarios.
All of these scenarios for $avoid$ need to be explicitly provided.
In this example, rule \eqref{eq:surpass_2} should be added as:
\begin{equation}
\begin{aligned}
\label{eq:prod_ex_2}
0.1:: & Avoid\_Player1\_Player2 & \rightarrow \\ & Waiting\_Player1\_Player2, &\\
& Running\_Player2 &
\end{aligned}
\end{equation} 
Note that now the probabilities of the above two production rules ($0.9$ and $0.1$) must sum to $1$.
The two scenarios are considered as mutually exclusive.
Note also that events are not relational.
On the other hand,
the addition of sensitivity in \cite{minnen_expectation_2003} and \cite{pei_learning_2013} 
can provide a more expressive power.

\newcolumntype{g}{>{\columncolor{Gray}}c}
\newcolumntype{H}{>{\setbox0=\hbox\bgroup}c<{\egroup}@{}}

%
%
\begin{table*}
\small
{
\resizebox{\textwidth}{!}
{
\begin{tabular}{lgcHgcgcgcgcHH} 
\toprule
\multicolumn{14}{c}{\bf Language Expressivity} \\
\midrule
{\bf Approach} &
{\bf $\sigma$} &  			
{\bf $\pi$} &     			
{\bf $\wedge$} & 			
{\bf $\vee$} &		
{\bf $\neg$} &
{\bf ;} &
{\bf *} &  		
{\bf W} &
{\bf Hierarchies} &
\pbox{3.5cm}{{\bf Temporal Model}} &
\pbox{1.6cm}{\centering {\bf Background Knowledge}} &
{\bf E.A.} &
{\bf Remarks} 
\\ 
\midrule		

%
%
\rowcolor{LightCyan}
\multicolumn{14}{c}{Automata} \\
\midrule

\parbox{3.5cm}{SASE+ \cite{kawashima_complex_2010}}
& 
{\checkmark} & {\checkmark} &  & 
 & & {\checkmark} & {\checkmark} & {\checkmark} & & Points,Implicit. &  & {\checkmark} &
 CEP
\\
\midrule

\parbox{3.5cm}{Lahar \cite{re_event_2008}}
& 
{\checkmark} & &  & 
 & & {\checkmark} & {\checkmark} & & & Points,Implicit. &   & {\checkmark} & CEP
\\
\midrule

\parbox{3.5cm}{Chuanfei et al. \citeyear{chuanfei_complex_2010}}
& 
 &  &  & 
 &  & {\checkmark} & {\checkmark} & {\checkmark} & & Points,Implicit. &  &
{\checkmark} &
\pbox{5.0cm}{CEP;\\ Not enough details in paper about $\sigma$, $\pi$, $\vee$, $\neg$.}
\\
\midrule

\parbox{3.5cm}{SASE+ AIG \cite{shen_probabilistic_2008}} 
& 
{\checkmark} & {\checkmark} & & 
 &  & {\checkmark} & {\checkmark} & {\checkmark} & & Points,Implicit. &  & {\checkmark} &
\pbox{5.0cm}{CEP;\\ Hierarchies supported but details unclear.}
\\
\midrule

\parbox{3.5cm}{Albanese et al. \citeyear{AlbaneseMPSU07,AlbaneseMPPS11}} 
& 
 & & & 
 &  & {\checkmark} & {\checkmark} & & & Points,Implicit. &  & & 
 \pbox{5.0cm}{Time constraints with upper bounds;\\
 $\sigma$ absent since event instances reified.} 
\\
\midrule

\parbox{3.5cm}{SASE+ optimized AIG \cite{wang_complex_2013}} 
& 
{\checkmark} & & & 
{\checkmark} & {\checkmark} & {\checkmark} & & {\checkmark} & {\checkmark} &
Points,Implicit. &  & {\checkmark} & CEP
\\
\midrule

\parbox{3.5cm}{SASE++ \cite{zhang_recognizing_2010,zhang_complexity_2014}}
& 
{\checkmark} & {\checkmark} & & 
{\checkmark} & {\checkmark} & {\checkmark} & {\checkmark} & {\checkmark} & &
Points,Implicit. & & {\checkmark} & CEP
\\
\midrule

%
%
\rowcolor{LightCyan}
\multicolumn{14}{c}{First-order logic \& Probabilistic Graphical Models} \\
\midrule

%
%
\parbox{3.5cm}{MLN-Allen \cite{morariu11}}
& 
{\checkmark} & {\checkmark} &  {\checkmark} & 
{\checkmark} & {\checkmark} & {\checkmark} & & {\checkmark} & {\checkmark} & 
\parbox{3.5cm}{\centering Intervals,Explicit. \\ Allen's Interval Algebra.} &
{\checkmark} & {\checkmark} &
\pbox{5.0cm}{+ Allen's Algebra predicates}
\\
\midrule

\parbox{3.5cm}{MLN-Event Calculus \\(Skarlatidis et al. \citeyear{skarlatidis_probabilistic_2011,skarlatidis_probabilistic_2015})} 
& 
{\checkmark} & {\checkmark} & {\checkmark} & 
{\checkmark} & {\checkmark} &  & & & {\checkmark} &
\parbox{3.5cm}{\centering Points,Explicit. \\ Event Calculus.}  & {\checkmark} &
{\checkmark} &
\pbox{5.0cm}{Event Calculus used}
\\
\midrule

\parbox{3.5cm}{MLN-hierarchical (Song et al. \citeyear{SongICMI13,SongSRAI13})}
& 
 &  &  & 
 {\checkmark} & {\checkmark} & {\checkmark} &  &  & {\checkmark} & 
 \parbox{3.5cm}{\centering Intervals,Explicit. \\ Allen's Interval Algebra.} & {\checkmark} &
\\
\midrule

\parbox{3.5cm}{ProbLog Event Calculus \cite{skarlatidis_probabilistic_2013}} 
& 
{\checkmark} & {\checkmark} & {\checkmark} & 
{\checkmark} & {\checkmark} &  & & & {\checkmark} & 
\pbox{3.5cm}{Points,Explicit. \\ Event Calculus.} & {\checkmark} &
{\checkmark} &
\pbox{5.0cm}{Event Calculus used}
\\
\midrule

\parbox{3.5cm}{Probabilistic Event Logic \cite{brendel2011PEL,selman2011PEL}} 
& 
 & & {\checkmark} & 
 {\checkmark} & {\checkmark} & {\checkmark} & 
 & & {\checkmark} & \parbox{3.5cm}{\centering Intervals,Implicit.\\ Allen's Interval Algebra.} & {\checkmark} & {\checkmark} & \pbox{5.0cm}{+
 Allen's Algebra operators}
\\
\midrule

\parbox{3.5cm}{Probabilistic Activity \\Detection \cite{AlbanesePADS10}} 
&
{\checkmark} & & {\checkmark} & 
 {\checkmark} & {\checkmark} & {\checkmark} & 
 & & & Intervals,Explicit. & & & \pbox{5.0cm}{
    $\sigma$ for time (in)equalities and object equalities; \\
    Arbitrarily nested quantifiers.
 }
\\
\midrule

\parbox{3.5cm}{KBMC Wasserkrug et al. \citeyear{wasserkrug2008complex,wasserkrug_model_2012,wasserkrug_efficient_2012}}
& 
{\checkmark} & {\checkmark} & {\checkmark} & 
{\checkmark} &  & {\checkmark} & & & {\checkmark} & Points,Explicit. & & {\checkmark} & CEP
\\
\midrule

\parbox{3.5cm}{CEP2U \cite{Cugola2014CEP2U}}
& 
{\checkmark} & {\checkmark} & {\checkmark} & 
 & {\checkmark} & {\checkmark} & & {\checkmark} & {\checkmark} & Points,Implicit. & & {\checkmark} & 
\pbox{5.0cm}{CEP; \\ * implicit.}
\\
\midrule

%
%
\rowcolor{LightCyan}
\multicolumn{14}{c}{Petri Nets} \\
\midrule

\parbox{3.5cm}{Probabilistic Petri Net \cite{AlbaneseCCMPSU08}} 
& 
 {\checkmark} & & {\checkmark} & 
{\checkmark} & {\checkmark} & {\checkmark} & 
{\checkmark} & & & Points,Implicit. &  & & 
\pbox{5.0cm}{$\sigma$ for temporal constraints}
\\
\midrule

\parbox{3.5cm}{Particle Filter Petri Net \cite{Lavee13PFPN}} 
& 
 & & {\checkmark} & 
{\checkmark} & & {\checkmark} & 
 & & & Points,Implicit. &  & &
\\
\midrule

%
%
\rowcolor{LightCyan}
\multicolumn{14}{c}{Grammars} \\
\midrule

\parbox{3.5cm}{\cite{ivanov_recognition_2000}} 
& 
 & & & 
{\checkmark} & & {\checkmark} & 
{\checkmark} & & {\checkmark} & Intervals,Implicit. &  & &
\\
\midrule

\parbox{3.5cm}{\cite{RyooA09}}
& 
{\checkmark} & & {\checkmark} & {\checkmark}
 & {\checkmark} & {\checkmark} & 
 & & {\checkmark} & 
 \parbox{3.5cm}{\centering Intervals,Implicit. \\ Allen's Interval Algebra.} &  & {\checkmark} & 
 \pbox{5.0cm}{
 + Allen's Algebra operators; \\
 + recursive definitions; \\
 $\sigma$ for spatial predicates; \\
 $\vee$, $\neg$ implicit.}
\\
\midrule

\parbox{3.5cm}{Temporal And-Or Graphs \cite{pei_learning_2013}} 
& 
{\checkmark} & & &
{\checkmark} & & {\checkmark} & 
 & & {\checkmark}  & 
 \parbox{3.5cm}{\centering Intervals,Implicit.} &  & {\checkmark} & 
 \pbox{5.0cm}{}
\\
\midrule

{\bf Approach} &
{\bf $\sigma$} &  			
{\bf $\pi$} &     			
{\bf $\wedge$} & 			
{\bf $\vee$} &		
{\bf $\neg$} &
{\bf ;} &
{\bf *} &  		
{\bf W} & 
{\bf Hierarchies} &
\pbox{3.5cm}{{\bf Temporal Model}} &
\pbox{1.6cm}{\centering {\bf Background Knowledge}} &
{\bf E.A.} &
{\bf Remarks} 
\\ 
\bottomrule

\end{tabular}
}
}
\caption{Expressive capabilities of CER systems. \newline
         $\sigma$: selection, $\pi$: production, 
		 $\vee$: disjunction, $\neg$: negation, 
		 ;: sequence, *: iteration, W: windowing}  
\label{table:lang}
\end{table*}

%
%
\begin{table*}
\small
{
\resizebox{\textwidth}{!}
{
\begin{tabular}{lgcHgcHgHHHH} 
\toprule
\multicolumn{10}{c}{\bf Probabilistic Expressivity} \\
\midrule
{\bf Approach} &
\pbox{1.5cm}{{\bf Model}} &
\pbox{2.2cm}{\centering {\bf Independence assumptions}} & 
\pbox{1.2cm}{{\bf Probability space}} &
\pbox{0.5cm}{\centering {\bf D.U.}} &
\pbox{0.5cm}{\centering {\bf P.U.}} &	
\pbox{1.6cm}{{\bf Generative/ Discriminative}} &
\pbox{0.5cm}{\centering {\bf H.C.}} & 
\pbox{1.0cm}{{\bf C. D.}} &
\pbox{0.8cm}{{\bf M. R.}} &
\pbox{1.5cm}{{\bf Remarks}} \\
\midrule		

%
%
\rowcolor{LightCyan}
\multicolumn{10}{c}{Automata} \\
\midrule

\parbox{3.5cm}{SASE+ \cite{kawashima_complex_2010}}
& 
Simple multiplication & 
\pbox{3.5cm}{All events independent.} &
{Events' history} & 
{Occ.} & 
& 
& 
& 
& 
& 

\\
\midrule

\parbox{3.5cm}{Lahar \cite{re_event_2008}}
&
Probabilistic Relational Algebra &
\parbox{4.5cm}{\centering 1st-order Markov for SDEs.\\ Different streams independent.} & 
{Events' history} & 
\pbox{1.6cm}{{Occ./ Att.}} &  
 & {Gener.} & & & &
\\
\midrule

\parbox{3.5cm}{Chuanfei et al. \citeyear{chuanfei_complex_2010}}
&
Simple multiplication &
\pbox{4.5cm}{1st-order Markov (+ some extensions).} & 
{? Events' history} & \pbox{1.6cm}{{Occ./ Att.}} &  & 
 {Gener.} & & & &
\\
\midrule

\parbox{3.5cm}{SASE+ AIG \cite{shen_probabilistic_2008}}
&
Lineage &
\pbox{3.5cm}{SDEs independent.} & 
{Events' history} & \pbox{1.6cm}{{Occ./ Att.}} &  & 
 {Gener.} & & & &
\\
\midrule

\parbox{3.5cm}{Albanese et al. \citeyear{AlbaneseMPSU07,AlbaneseMPPS11}} 
&
\parbox{4.5cm}{\centering Patterns modeled as stochastic processes, similar to Markov chains.} &
\parbox{4.5cm}{\centering 1st-order Markov within CE. Different CEs independent.} &  
{? Events' history} & & {\checkmark} & 
{Gener.} 
& & & {\checkmark} &
\pbox{3.5cm}{Patterns modeled as stochastic activities/processes, similar to Markov chains.}
\\
\midrule

\parbox{3.5cm}{SASE+ optimized AIG \cite{wang_complex_2013}}
& 
Multiplication on Markov chain & 
\parbox{4.5cm}{\centering SDEs independent or Markovian.\\ Different streams independent.} &
{Events' history} & {Occ.} & & 
 {Gener.} & & & &
\\
\midrule

\parbox{3.5cm}{SASE++ \cite{zhang_recognizing_2010,zhang_complexity_2014}} 
& 
\parbox{4.5cm}{\centering Probability distribution on time attribute.} &
{SDEs independent.} & 
{Events' history} & Occ. & & 
{Gener.} & & & &
\pbox{4.5cm}{Probability distribution on time attribute.}
\\
\midrule

%
%
\rowcolor{LightCyan}
\multicolumn{10}{c}{First-order logic \& Probabilistic Graphical Models} \\
\midrule

%
%

\parbox{3.5cm}{MLN-Allen \cite{morariu11}}
& 
\parbox{4.5cm}{\centering Markov Logic Networks.\\
 Bottom-up hypothesis generation.} &
 None. &
Interpretations & Occ. & {\checkmark} & 
  Gener.  & {\checkmark} & & & 
 \parbox{4.5cm}{\centering Markov Logic Networks.\\
 Bottom-up hypothesis generation.}
\\
\midrule

\parbox{3.5cm}{MLN-Event Calculus (Skarlatidis et al. \citeyear{skarlatidis_probabilistic_2011,skarlatidis_probabilistic_2015})} 
& 
\pbox{4.5cm}{Markov Logic Networks.} &
None. &
Interpretations &  & {\checkmark} & 
 Gener. & {\checkmark} & & & 
\pbox{3.5cm}{Markov Logic Networks.}
\\
\midrule

\parbox{3.5cm}{MLN-hierarchical (Song et al. \citeyear{SongICMI13,SongSRAI13})}
& 
\pbox{4.5cm}{Markov Logic Networks.} &
None  &  
&  & 
{\checkmark}  &   & {\checkmark} & & 
\\
\midrule

\parbox{3.5cm}{ProbLog Event Calculus \cite{skarlatidis_probabilistic_2013}} 
& 
\pbox{3.5cm}{Probabilistic Logic Programming.} &
SDEs independent. &
Proofs(?) & Occ. & & 
 Gener. & & & {\checkmark} & \pbox{3.5cm}{ProbLog}
\\
\midrule

\parbox{3.5cm}{Probabilistic Event Logic \cite{brendel2011PEL,selman2011PEL}} 
& 
\parbox{3.5cm}{\centering Weights, as in Conditional Random Fields.} &
None &
Interpretations & & {\checkmark} & 
 Discr. & {\checkmark} & & & 
\pbox{3.5cm}{Weights, as in Conditional Random Fields.\\ Weights learned.}
\\
\midrule

\parbox{3.5cm}{Probabilistic Activity \\Detection \cite{AlbanesePADS10}} 
&
\parbox{4.5cm}{\centering Probabilities assigned to predicates for object equality.} &
\pbox{3.5cm}{Depends on t-norm.} & 
Eqs upon objects & Occ. & & 
 ? &  & & & 
\parbox{3.5cm}{\centering Probabilities assigned to predicates for object equality.}
\\
\midrule

\parbox{3.5cm}{KBMC Wasserkrug et al. \citeyear{wasserkrug2008complex,wasserkrug_model_2012,wasserkrug_efficient_2012}}
& 
\pbox{3.5cm}{Bayesian Networks.} &
{SDEs independent.} &
{Events' history} & \pbox{1.6cm}{{Occ./ Att.}} & {\checkmark} & 
 {Gener.} & & & & 
\pbox{3.5cm}{Bayesian Networks.}
\\
\midrule

\parbox{3.5cm}{CEP2U \cite{Cugola2014CEP2U}}
&
\pbox{3.5cm}{Bayesian Networks.} &
\parbox{4.5cm}{\centering Event attributes independent. \\ SDEs independent. \\
CEs dependent only on events immediately below in hierarchy.} & 
? & \pbox{1.6cm}{{Occ./ Att.}} & {\checkmark} & 
 {Gener.} & & {\checkmark} & & 
\pbox{3.5cm}{Bayesian Networks}
\\
\midrule

%
%
\rowcolor{LightCyan}
\multicolumn{10}{c}{Petri Nets} \\
\midrule

\parbox{3.5cm}{Probabilistic Petri Net \cite{AlbaneseCCMPSU08}} 
& 
\parbox{4.5cm}{\centering Hard constraints as forbidden actions.\\
Activities as stochastic processes.}
&
\parbox{4.5cm}{\centering Conditioned on previous event in CE pattern.} &
Events' history? & & {\checkmark} & 
 Gener. & {\checkmark} & & & 
\pbox{3.5cm}{Hard constraints as forbidden actions;\\
As in \cite{AlbaneseMPPS11}, activities resemble stochastic processes.}
\\
\midrule

\parbox{3.5cm}{Particle Filter Petri Net \cite{Lavee13PFPN}} 
& 
Bayesian recursive filter &
\parbox{4.5cm}{\centering 1st-order Markov. \\ SDEs independent.} &
? & Occ. & & 
 Gener. & & & & 
\\
\midrule

%
%
\rowcolor{LightCyan}
\multicolumn{10}{c}{Grammars} \\
\midrule

\parbox{3.5cm}{\cite{ivanov_recognition_2000}} 
&
\parbox{4.5cm}{\centering Stochastic production rules.} &
\pbox{4.5cm}{Rules conditionally independent.} & 
? & Occ. & {\checkmark} & 
 Gener.?& & & {\checkmark} & 
\pbox{3.5cm}{Stochastic production rules.} 
\\
\midrule

\parbox{3.5cm}{\cite{RyooA09}}
& 
\pbox{4.5cm}{Similar to Bayesian Networks \\(but siblings not cond. independent)}&
\pbox{4.5cm}{Conditional independence of SDEs.} &
 & Occ. &  & 
 Gener. & & & & 
\\
\midrule

\parbox{3.5cm}{Temporal And-Or Graphs \cite{pei_learning_2013}} 
& 
\pbox{4.5cm}{As in Markov Chains for Or-nodes.\\ As in graphical models for And-nodes.} &
 None. &
 & Occ. & {\checkmark} & 
 & & & & 
\\
\midrule

{\bf Approach} &
\pbox{1.5cm}{{\bf Model}} &
\pbox{2.2cm}{\centering {\bf Independence assumptions}} & 
\pbox{1.2cm}{{\bf Probability space}} &
\pbox{0.5cm}{\centering {\bf D.U.}} &
\pbox{0.5cm}{\centering {\bf P.U.}} &	
\pbox{1.6cm}{{\bf Generative/ Discriminative}} & 
\pbox{0.5cm}{\centering {\bf H.C.}} & 
\pbox{1.0cm}{{\bf C.D.}} &
\pbox{0.8cm}{{\bf M.R.}} &
\pbox{1.5cm}{{\bf Remarks}} \\

\bottomrule

\end{tabular}
}
}
\caption{Expressive power of CER systems with respect to their probabilistic properties. 
D.U.: Data Uncertainty, Occ.:Occurrence, Att.:Attributes, P.U.: Pattern Uncertainty, H.C.: Hard Constraints}
\label{table:prob}
\end{table*}

%
%

\begin{table*}
\small
{
\resizebox{\textwidth}{!}
{
\begin{tabular}{lHgcgcgH}
\toprule
\multicolumn{7}{c}{\bf Inference} \\
\midrule
{\bf Approach} & 
{\bf Queries' domain} &  
\pbox{1.5cm}{{\bf Type}} & 
\pbox{1.5cm}{{\bf Confidence Thresholds}} & 
{\bf Approx.} & 
\pbox{1.5cm}{{\bf Distrib.}} &
{\bf Performance} & 
{\bf Remarks}  \\  			 					
\midrule		

%
%
\rowcolor{LightCyan}
\multicolumn{7}{c}{Automata} \\
\midrule

\parbox{3.5cm}{SASE+ \cite{kawashima_complex_2010}}
& 
{Events} & {Marginal} & 
{\checkmark} & & &  
\parbox{6.0cm}{\centering 0.8-1.1 K events/sec with $Kleene +$.} & 
\\
\midrule

\parbox{3.5cm}{Lahar \cite{re_event_2008}}
& 
{Events} & {Marginal} & {\checkmark} 
 & & &
 \parbox{6.0cm}{\centering $> 10$ points increase in accuracy.\\
~ 100K events/sec for Extended Regular Queries.} & 
\\
\midrule

\parbox{3.5cm}{Chuanfei et al. \citeyear{chuanfei_complex_2010}}
& 
{Events} & {Marginal} & {\checkmark} & & &
\parbox{6.0cm}{\centering 4-8K events/sec for patterns of length 6 down to 2.} 
 & 
\\
\midrule

\parbox{3.5cm}{SASE+ AIG \cite{shen_probabilistic_2008}} 
& 
{Events} & {Marginal} & {\checkmark}
 & & &
\parbox{6.0cm}{\centering 1000K events/sec, almost constant for varying window size (3 to 15 timepoints).\\
1000K-100K events/sec for experiments with 1 up to 10 alternatives of a SDE.}
& 
\\
\midrule

\parbox{3.5cm}{Albanese et al. \citeyear{AlbaneseMPSU07,AlbaneseMPPS11}} 
& 
{Events} & {Marginal and MAP} & 
{\checkmark} & & {\checkmark} &
\parbox{6.0cm}{\centering Running time linear in video length.\\ 
Parallel version reached 335K events/sec with 162 computing nodes.} & 
\pbox{3.0cm}{Can detect unexplained activities. \\
Parallel version in \cite{MolinaroMPPRS14}.}
\\
\midrule

\parbox{3.5cm}{SASE+ optimized AIG \cite{wang_complex_2013}}
& 
{Events} & {Marginal} & 
{\checkmark} & & {\checkmark} &
\parbox{6.0cm}{\centering 8K-13K events/sec for 2-6 nodes.}
& Distributed
\\
\midrule

\parbox{3.5cm}{SASE++ \cite{zhang_recognizing_2010,zhang_complexity_2014}}
& 
{Events} & {Marginal} & {\checkmark}
 & & &
 \parbox{6.0cm}{\centering Reduction from exponential to close-linear cost w.r.t to selectivity / window size.}
 & 
\\
\midrule

%
%
\rowcolor{LightCyan}
\multicolumn{7}{c}{First-order logic \& Probabilistic Graphical Models} \\
\midrule

%
%

\parbox{3.5cm}{MLN-Allen \cite{morariu11}}
& 
 & Marginal &  & &  &
 \parbox{6.0cm}{\centering F-measure $>$ 70\% for varying window sizes.} &   
\\
\midrule

\parbox{3.5cm}{MLN-Event Calculus (Skarlatidis et al. \citeyear{skarlatidis_probabilistic_2011,skarlatidis_probabilistic_2015})} 
& 
 & Marginal &  & {\checkmark} &  &
 \parbox{6.0cm}{\centering Increased precision, slight decrease in recall, compared to deterministic solution.} &   
\\
\midrule

\parbox{3.5cm}{MLN-hierarchical (Song et al. \citeyear{SongICMI13,SongSRAI13})}
& 
 & MAP &  &  &
  &    &
\\
\midrule

\parbox{3.5cm}{ProbLog Event Calculus \cite{skarlatidis_probabilistic_2013}} 
& 
 & Marginal &  & &  &
 \parbox{6.0cm}{\centering Improved F-measure w.r.t. crisp version.} &   
\\
\midrule

\parbox{3.5cm}{Probabilistic Event Logic \cite{brendel2011PEL,selman2011PEL}} 
& 
 & MAP & & {\checkmark} &  &
 \parbox{6.0cm}{\centering Smooth accuracy degradation when noise in time intervals of SDEs added. \\ 
 Relative robustness against false positives/negatives.} & 
 
\\
\midrule

\parbox{3.5cm}{Probabilistic Activity \\Detection \cite{AlbanesePADS10}} 
& 
\pbox{1.2cm}{Events/ Minimal subevents} & Marginal & {\checkmark}  & &  &
\parbox{6.0cm}{\centering Running time at most linear in the number of atoms. \\ 
Better precision/recall than Hidden Markov Models/ Dynamic Bayesian Networks, higher computation time.} &
\pbox{3.0cm}{2 versions: On-/Off-line}
\\
\midrule

\parbox{3.5cm}{KBMC Wasserkrug et al. \citeyear{wasserkrug2008complex,wasserkrug_model_2012,wasserkrug_efficient_2012}}
& 
{Events} & {Marginal} & 
 & {\checkmark} &  &
 \parbox{6.0cm}{\centering CEs within desired confidence interval. \\
Sub-linear decay of event rate w.r.t possible worlds.} 
 & 
\\
\midrule

\parbox{3.5cm}{CEP2U \cite{Cugola2014CEP2U}}
& 
{Events} & {Marginal} & 
{\checkmark} &  &  &
\parbox{6.0cm}{\centering 50\% overhead w.r.t deterministic case.}
&
\\
\midrule

%
%
\rowcolor{LightCyan}
\multicolumn{7}{c}{Petri Nets} \\
\midrule

\parbox{3.5cm}{Probabilistic Petri Net \cite{AlbaneseCCMPSU08}} 
& 
 & {Marginal and MAP} & {\checkmark} & & & 
 \parbox{6.0cm}{\centering $\sim$ 3 seconds to process videos $\sim$ with 60 different SDE types.}
 & 
\\
\midrule

\parbox{3.5cm}{Particle Filter Petri Net \cite{Lavee13PFPN}} 
& 
 & Marginal & & {\checkmark} & & 
 \parbox{6.0cm}{\centering Increased true positive rate, compared to deterministic solution. Slight increase in false positive rate.} & 
\\
\midrule

%
%
\rowcolor{LightCyan}
\multicolumn{7}{c}{Grammars} \\
\midrule

\parbox{3.5cm}{\cite{ivanov_recognition_2000}} 
& 
 & Marginal & {\checkmark} & & &  &
\\
\midrule

\parbox{3.5cm}{\cite{RyooA09}}
& 
 & Marginal & {\checkmark} & &  &
 \parbox{6.0cm}{\centering Increased accuracy when noisy SDEs present, compared against case with crisp SDEs.} & 
\\
\midrule

\parbox{3.5cm}{Temporal And-Or Graphs \cite{pei_learning_2013}}
& 
 & MAP & {\checkmark} & &  &
 \parbox{6.0cm}{\centering 87\%--90\% accuracy for predicting human intentions.} & 
\\
\midrule

{\bf Approach} & 
{\bf Queries' domain} &  
\pbox{1.5cm}{{\bf Type}} & 
\pbox{1.5cm}{{\bf Confidence Thresholds}} & 
{\bf Approx.} & 
\pbox{1.5cm}{{\bf Distrib.}} &
{\bf Performance} & 
{\bf Remarks}  \\
\bottomrule

\end{tabular}
}
}
\caption{Inference capabilities of probabilistic CER systems}
\label{table:inf}
\end{table*}

\section{Discussion}
\label{section:conclusions}

\newcommand{\tabitem}{~~\llap{\textbullet}~~}
\begin{table*}
\small
\resizebox{\textwidth}{!}
{
\begin{tabular}{lll} 
\toprule
{\bf Approach} &
{\bf Strengths} &
{\bf Weaknesses}  
\\ 
\midrule

Automata 
&
\pbox{5.0cm}{\textcolor{white}{dummy} \\ $Iteration$, $\mathit{Windowing}$, formal Event Algebra.
\vspace{3mm}
\\ Data uncertainty, both with respect to occurrence of events
				      and event attributes. 
\vspace{3mm}				      
\\ Support for confidence thresholds. High throughput values.			
} 
& 
\pbox{5.0cm}{\textcolor{white}{dummy} \\ Limited support for event hierarchies. No background knowledge.
Implicit time representation (hence no explicit constraints on time attribute).
\vspace{3mm}
\\ Limited or no support for rule uncertainty. Too many independence assumptions.
No hard constraints. 
\vspace{3mm}
\\
Throughput figures come from experiments with simplistic event patterns.} 
\\
\midrule

\pbox{3.0cm}{First-order logic \& Probabilistic Graphical Models}
&
\pbox{5.0cm}{\textcolor{white}{dummy} \\   Complex temporal patterns, with explicit time constraints.
				      Event hierarchies.
				      Background knowledge.
				      Usually provide a formal Event Algebra.
\vspace{3mm}
\\
				      Pattern uncertainty.
				      Limited independence assumptions.
				      Hard constraints possible.
\vspace{3mm}
\\
				      MAP and approximate inference.
}
 & 
\pbox{5.0cm}{\textcolor{white}{dummy} \\ No $\mathit{Iteration}$. 
				      Limited support for $\mathit{Windowing}$.
\vspace{3mm}
\\
				 Harder (but not impossible) to express data uncertainty.
				      Often training required (experts cannot ``simply'' assign probabilities to rules).
\vspace{3mm}
\\
				 Low (or unknown) throughput.}
 
\\
\midrule

Petri Nets 
&
\pbox{5.0cm}{\textcolor{white}{dummy} \\  Concurrency and synchronization.
\vspace{3mm}
\\
			  Support for both data and pattern uncertainty (but not both in the same model).
\vspace{3mm}
\\
			  Can perform both MAP and Marginal inference.
		      Confidence thresholds and approximate inference possible.}
&
\pbox{5.0cm}{\textcolor{white}{dummy} \\ Not truly relational.
				      No $\mathit{Windowing}$, hierarchies or background knowledge.
				      Implicit time representation.
\vspace{3mm}
\\
				      Strict independence assumptions.
\vspace{3mm}
\\
				      Low (or unknown) throughput.} 
 
\\
\midrule

Grammars 
&
\pbox{5.0cm}{\textcolor{white}{dummy} \\  Very easy to model hierarchies and $Iteration$. Recursive patterns.
\vspace{3mm}
\\
				 	 Both data and pattern uncertainty.
\vspace{3mm}
\\
					 Confidence thresholds.}
&
\pbox{5.0cm}{\textcolor{white}{dummy} \\ Not truly relational (unless context-sensitive grammars are used).
				      No $\mathit{Negation}$.
				      Implicit time representation.
				      No background knowledge.
\vspace{3mm}
\\
					  No rich methodology for efficient probabilistic inference.
\vspace{3mm}
\\
					  Unknown performance for throughput.
} 
 
\\

\bottomrule

\end{tabular}
}
\caption{Strengths and weaknesses of the reviewed probabilistic CER approaches}
\label{table:conclusions}
\end{table*}

Table \ref{table:lang} summarizes the expressive power of the presented CER systems.
Its first columns correspond to the list of operators presented in Section \ref{section:eval:rep}.
The other columns assess various aspects of the functionality supported by each
system.
These are:
\begin{itemize}
  \item Hierarchies: The ability to define CEs at
           various levels and reuse those intermediate inferred events in order
           to infer other higher-level events.
  \item Temporal Model.: Events may be represented by timepoints
  			  (P) or intervals (In). Moreover, the time attribute might be explicitly
  			  included in the constraints (Ex) (e.g. $(T_{2}>T_{1}) \wedge (T_{2}-T_{1}>100)$) 
  			  or temporal constraints may be defined by
  			  referring implicitly to time in the rules (Im) 
  			  (e.g. the $Sequence$ operator implicitly defines $T_{2}>T_{1}$).
  \item Background Knowledge.: Does the system support knowledge,
  			  besides the CE patterns?
\end{itemize}

In Table \ref{table:prob} we present the probabilistic properties of each method:
\begin{itemize}
  \item Model: The probabilistic model used.
  \item Independence assumptions: What are the independence/dependence
        assumptions made? Note that when we write that a method makes no independence assumptions
        (cell filled with ``None''), the meaning is not that it cannot make such assumptions,
        but that it does not have to and that it does not enforce them on the probabilistic model.
  \item Data uncertainty: Does the system support data uncertainty? If yes, is it about
        the \textit{occurrence} of events or \textit{both} about the occurrence
        and the attributes?
  \item Pattern uncertainty: Is there support for uncertain patterns?
  \item Hard constraints: Is there support for rules that may not be violated?
\end{itemize}

Table \ref{table:inf} presents the inference capabilities of the presented systems along the following lines:
\begin{itemize}
  \item Type: Can the system perform \textit{Marginal} inference,
        \textit{MAP} inference or \textit{both}?
  \item Confidence Thresholds: Is there support support for confidence thresholds above
        which a CE is accepted?
  \item Approximate: Does the system support techniques for approximate
        inference?
  \item Distribution: Is there a distributed version of the proposed solution? 
  \item Performance: Remarks about performance with respect to throughput and accuracy.
	Note that such information is not always available.
        Moreover, since standard benchmarks for probabilistic CER are not available,
        the presented figures are those reported by the authors of each approach,
        who may choose metrics and datasets for their own purposes.
        This means that a direct comparison is not possible.
        We are intending to look into the possibility of such an empirical evaluation as part of our future work.
\end{itemize}

Our review of probabilistic CER systems identified a number of strengths and limitations
for the proposed approaches.
We summarize our conclusions in Table \ref{table:conclusions}.
As a note of caution though,
we note that, when a weakness is reported,
this does not mean that the corresponding method cannot in general support a feature
(e.g., as might be the case for $\mathit{Iteration}$ in first-order logic),
but that the presented methods have not incorporated it,
although it might be possible 
(for example, the absence of hierarchies in automata-based methods).

The current systems for probabilistic CER need to deal with a trade-off between 
language expressivity, probabilistic expressivity and complexity. 
As can be seen in Table \ref{table:conclusions},
automata-based methods can easily handle $sequence$ and $iteration$ operators,
but they usually model only data uncertainty,
without taking into account pattern uncertainty.
Moreover, they rarely move beyond the 1st-order Markov assumption.
Therefore, when more expressive power is required,
one of the other three approaches should be preferred.
For example, Petri Nets are quite powerful for modeling concurrency and synchronization.
Grammars are very well suited for expressing hierarchies and recursive patterns 
and can also model both data and pattern uncertainty.
If a truly relational model is needed, 
than first-order logic approaches can provide a solution
and can also handle intervals, hierarchies and background knowledge.
When combined with probabilistic graphical models,
they can also be very flexible as far as their independence assumptions are concerned.
However, this increased expressive power comes at the cost of high computational complexity
and, as a result, of low throughput.

Achieving high throughput figures with more complex patterns/models so that real-time inference
is possible still remains a challenge
and very few approaches have explored distributed solutions.
The performance trade-offs between throughput and latency have also not been explored yet.
Moreover, only rarely do these systems attempt to learn the weights or structure of patterns and
instead rely on experts to define them.
This might be sufficient for simple patterns,
but would not scale well in case multiple, complex rules with complicated dependencies are required.
Finally, in order to assess the capabilities of probabilistic CER systems,
there is a need for standard benchmarks that would allow for empirical comparisons in various scenarios.

\section*{Acknowledgement}
This work has been funded by the EU SPEEDD project (FP7-ICT 619435).

\bibliographystyle{plainnat}
\bibliography{ms}
\end{document}